\documentclass[pre,tighten,eqsecnum,showpacs]{revtex4}

\usepackage{amssymb,amsmath}

\usepackage{graphicx}

\begin{document}
\title{Shear-rate dependent transport coefficients for inelastic Maxwell models}
\author{Vicente Garz\'{o}}
\email{vicenteg@unex.es} \homepage{http://www.unex.es/eweb/fisteor/vicente/}
\address{Departamento de F\'{\i}sica, Universidad de Extremadura, E-06071
Badajoz, Spain}
 \begin{abstract}
The Boltzmann equation for $d$-dimensional inelastic Maxwell models is considered to analyze transport
properties in spatially inhomogeneous states close to the simple shear flow. A normal solution is obtained via a
Chapman--Enskog--like expansion around a local shear flow distribution $f^{(0)}$ that retains all the
hydrodynamic orders in the shear rate. The constitutive equations for the heat and momentum fluxes are obtained
to first order in the deviations of the hydrodynamic field gradients from their values in the reference state
and the corresponding generalized transport coefficients are {\em exactly} determined in terms of the
coefficient of restitution $\alpha$ and the shear rate $a$. Since $f^{(0)}$ applies for arbitrary values of the
shear rate and is not restricted to weak dissipation, the transport coefficients turn out to be nonlinear
functions of both parameters $a$ and $\alpha$. A comparison with previous results obtained for inelastic hard
spheres from a kinetic model of the Boltzmann equation is also carried out.


\end{abstract}

\pacs{05.20.Dd, 45.70.-n, 51.10.+y}
\draft
\date{\today}
\maketitle

\section{Introduction}
\label{sec1}

Granular gases are usually modelled as a gas of hard spheres dissipating part of their kinetic energy during
collisions. In the simplest model, the grains are taken to be smooth so that the inelasticity of collisions is
characterized only through a constant coefficient of normal restitution $\alpha \leq 1$. For a low-density gas,
the Boltzmann kinetic equation has been conveniently modified to account for inelastic binary collisions
\cite{GS95,BP04} and the Navier--Stokes transport coefficients \cite{single} for states with small hydrodynamic
gradients have been computed by means of the Chapman--Enskog method \cite{CC70} around the local version of the
homogeneous cooling state. As in the case of elastic collisions, the transport coefficients are given in terms
of the solutions of exact linear integral equations which are approximately solved by considering the leading
terms in a Sonine polynomial expansion of the velocity distribution function. On the other hand, in spite of
this approach, the Sonine predictions compare in general quite well with computer simulations, even for quite
strong dissipation conditions \cite{computer}.

Needless to say, the mathematical difficulties of solving the Boltzmann equation for inelastic hard spheres
(IHS) increase considerably when one considers situations for which large gradients occur and more complex
constitutive equations than the Navier--Stokes ones (which  are linear in the spatial gradients) are required.
For this kind of situations one has to resort to alternative approaches, such as the use of simplified kinetic
models, where the true Boltzmann collision operator is replaced by a simpler collision model (e.g., the BGK
model \cite{C88}) that preserves its most relevant physical features. In the case of elastic collisions, this
route has been shown to be very fruitful since most of the exact solutions derived from the BGK model agree
rather well \cite{GS03} with results obtained by numerically solving the Boltzmann equation by means of the
direct simulation Monte Carlo (DSMC) method \cite{Bird}. Nevertheless, much less is known for inelastic
collisions although some exact results obtained in the simple or uniform shear flow (USF) problem \cite{USF} and
in the nonlinear Couette flow state \cite{TTMGSD01} support the reliability of the inelastic version of the BGK
model \cite{USF} for granular gases as well. The USF state is perhaps the simplest flow problem since the only
nonzero hydrodynamic gradient is $\partial u_x/\partial y\equiv a=\text{const}$, where ${\bf u}$ is the flow
velocity and $a$ is the constant shear rate. Due to its simplicity, this state has been widely studied in the
past for elastic \cite{GS03} and inelastic gases \cite{C90} as an ideal testing ground to shed light on the
intricacies associated with the response of the system to strong shear rates. Very recently \cite{G06}, the
inelastic BGK model has been applied to determine the transport coefficients for a granular gas in
spatially-inhomogeneous states close to the USF. The heat and momentum fluxes were evaluated to first order in
the deviations of the hydrodynamic field gradients from their values in the reference USF state. Given that the
system is strongly sheared, the corresponding transport coefficients turn out to be {\em nonlinear} functions of
both the shear rate and the coefficient of restitution. This is the main new ingredient of these constitutive
equations. In addition, due to the mathematical difficulties involved in the general problem, results were
restricted to a particular kind of perturbations for which the steady state conditions of the USF hold
\cite{G06}. This allowed us to perform a linear stability analysis of the hydrodynamic equations with respect to
the USF state.

As happens for elastic collisions, a possibility to analyze transport around USF retaining the explicit form of
the Boltzmann collision operator is to consider inelastic Maxwell models (IMM), i.e., models for which the
collision rate is independent of the relative velocity of the two colliding particles. Although these IMM do not
correspond to any microscopic interaction, it has been shown by several authors \cite{IMM} that the cost of
sacrificing physical realism can be in part compensated by the amount of exact analytical results derived from
this interaction model. This is the main reason why IMM have attracted the attention of physicists and
mathematicians since the beginning of the century. On the other hand, beyond its academic interest, it must be
remarked that recent experiments \cite{KSSAON05} for magnetic grains with dipolar interactions in air in a
two-dimensional geometry have been accurately described by IMM. The excellent quantitative agreement found
between the magnetic particles experiments \cite{KSSAON05} and IMM suggests that magnetic particles can be
considered as an ideal experimental probe for the predictions of this analytically tractable kinetic theory.

The main advantage of using IMM instead of IHS is that a velocity moment of order $k$ of the Boltzmann collision
operator only involves moments of order less than or equal to $k$. This allows one to evaluate the Boltzmann
collision moments without the explicit knowledge of the velocity distribution function \cite{GS07}. Thanks to
this simplification, the velocity moments of IMM under USF can be computed as functions of the shear rate. Very
recently \cite{SG06}, the first nontrivial shear-rate dependent moments of USF for IMM have been explicitly
computed: the second-degree moments which are directly related to rheological properties and the fourth-degree
moments whose knowledge is necessary to get the heat flux around USF. The knowledge of these moments allows one
to reexamine the problem studied in Ref.\ \cite{G06} in the context of the Boltzmann equation and without any
restriction on the kind of perturbations to USF considered. In addition, the comparison between the transport
coefficients derived here with those obtained from IHS \cite{G06} can be used again as a test to assess the
degree of usefulness of IMM as a prototype model for granular flows. Previous comparisons between IMM and IHS
for inhomogeneous situations \cite{S03,G03,GA05} have shown in general good agreement (especially for low order
moments), confirming the reliability of IMM as a good approximation to model granular systems.

Since I am interested in studying heat and momentum transport in a strongly sheared granular gas, the physical
situation is such that the gas is in a state that deviates from the USF by {\em small} spatial gradients. Under
these conditions and taking the USF state $f^{(0)}$ as the reference one, the Boltzmann equation is solved by
means of a Chapman--Enskog-like expansion \cite{L06} around the distribution $f^{(0)}$. Since the latter applies
for arbitrary values of the shear rate $a$, the successive approximations in the Chapman--Enskog expansion will
retain all the hydrodynamic orders in $a$. Therefore, the non-equilibrium problem studied here deals with two
kind of spatial gradients: {\em small} gradients due to perturbations of the USF and arbitrary {\em large}
gradients due to the background shear flow. In this paper, the calculations will be restricted to first order in
the spatial gradients of density, temperature and flow velocity. At this level of approximation, the momentum
transport is characterized by a viscosity tensor $\eta_{ijk\ell}$, while the heat flux is expressed in terms of
a thermal conductivity tensor $\kappa_{ij}$ and a new tensor $\mu_{ij}$ (not present in the elastic case). The
set of generalized transport coefficients $\eta_{ijk\ell}$, $\kappa_{ij}$, and $\mu_{ij}$ are nonlinear
functions of $a$ and $\alpha$. The determination of this dependence for IMM is the primary target of this paper.

The plan of the paper is as follows. In Sec.\ \ref{sec2}, the Boltzmann equation for IMM is introduced and a
brief summary of relevant results derived for the USF problem is given. The Chapman--Enskog expansion around USF
is described in Sec.\ \ref{sec3} and the linear integral equations defining the generalized transport
coefficients are displayed. Section \ref{sec4} deals with the explicit evaluation of the transport coefficients
associated with the momentum and heat fluxes. The details of the calculations are displayed along several
Appendices. The dependence of some of the above coefficients on the shear rate and on the coefficient of
restitution is illustrated and compared with known results obtained for IHS \cite{G06}. The comparison shows in
general qualitative good agreement, especially in the case of the viscosity tensor $\eta_{ijk\ell}$.  The paper
is closed in Sec.\ \ref{sec5} with a brief discussion on the results reported in this paper.

\section{Inelastic Maxwell models and uniform shear flow}
\label{sec2}

Let us consider a granular fluid modelled as an inelastic Maxwell gas. The inelasticity of collisions among all
pairs is accounted for by a {\em constant} coefficient of restitution $0\leq \alpha \leq 1$ that only affects
the translational degrees of freedom of the grains. At a kinetic level, all the relevant information on the
state of the system is provided by the one-particle velocity distribution function $f({\bf r}, {\bf v}, t)$. In
the low density regime the {\em inelastic} Boltzmann equation \cite{GS95,BP04} gives the time evolution of
$f({\bf r}, {\bf v}, t)$. The corresponding Boltzmann equation for inelastic Maxwell models (IMM) can be
obtained from the inelastic Boltzmann equation for inelastic hard spheres (IHS) by replacing the rate for
collisions between two particles (which is proportional to the relative velocity in the case of IHS) by an
average velocity-independent collision rate. With this simplification and in the absence of an external force,
the Boltzmann equation for IMM reads
\begin{equation}
\label{2.1} \left(\frac{\partial}{\partial t}+{\bf v}\cdot \nabla
\right)f({\bf r}, {\bf v},t)=J[{\bf v}|f(t),f(t)],
\end{equation}
where the Boltzmann collision operator is
\begin{equation}
\label{2.2} J\left[{\bf v}_{1}|f,f\right] =\frac{\omega}{n\Omega_d} \int d{\bf v}_{2}\int
d\widehat{\boldsymbol{\sigma}} \left[ \alpha^{-1}f({\bf v}_{1}')f({\bf v}_{2}')-f({\bf v}_{1})f({\bf
v}_{2})\right] \;.
\end{equation}
Here, $n$ is the number density, $\Omega_d=2\pi^{d/2}/\Gamma(d/2)$
is the total solid angle in $d$ dimensions,
$\widehat{\boldsymbol{\sigma}}$ is a unit vector along the line of
the two colliding spheres, and ${\bf g}={\bf v}_1-{\bf v}_2$ is
the relative velocity of the colliding pair. In addition, the
primes on the velocities denote the initial values $\{{\bf
v}_{1}^{\prime}, {\bf v}_{2}^{\prime}\}$ that lead to $\{{\bf
v}_{1},{\bf v}_{2}\}$ following a binary collision:
\begin{equation}
\label{2.3} {\bf v}_{1}^{\prime}={\bf v}_{1}-\frac{1}{2}\left(
1+\alpha ^{-1}\right)(\widehat{\boldsymbol{\sigma}}\cdot {\bf
g})\widehat{\boldsymbol {\sigma}}, \quad {\bf v}_{2}^{\prime}={\bf
v}_{2}+\frac{1}{2}\left( 1+\alpha^{-1}\right)
(\widehat{\boldsymbol{\sigma}}\cdot {\bf
g})\widehat{\boldsymbol{\sigma}}\;.
\end{equation}
The effective collision frequency $\omega$ is independent of velocity but depends on space and time through its
dependence on density and temperature. Here, I will assume that $\omega \propto n T^q$, with $q \geq 0$. The
case $q=0$ will be referred here as Model A, while the case $q \neq 0$ will be called Model B. Model A is closer
to the original Maxwell model of elastic particles, while model B is closer to hard spheres when
$q=\frac{1}{2}$. Furthermore, the collision frequency $\omega$ can be also seen as a free parameter of the
model, determined to optimize the agreement with some property of interest of the original Boltzmann equation
for IHS. As in previous works \cite{S03,G03,GA05}, $\omega$ is chosen here to guarantee that the cooling rate of
IMM be the same as that of IHS (evaluated at the local equilibrium approximation)\cite{S03}. With this choice,
one gets
\begin{equation}
\label{2.14} \omega=\frac{d+2}{2}\nu_0,\quad \nu_0=A(q)n T^q,
\end{equation}
where the value of the quantity $A(q)$ is irrelevant for our purposes. Henceforth, I will take this choice for
$\omega$.

There is another more refined version of IMM \cite{BCG00} where the collision rate  has the same dependence on
the scalar product ($ \widehat{\boldsymbol{\sigma }}\cdot \widehat{{\bf g}}$) as in the case of IHS. However,
both versions of IMM lead to similar results in problems as delicate as the high energy tails and so I will
consider here the simplest version given by Eqs.\ (\ref{2.2}) and (\ref{2.3}).

The first $d+2$ velocity moments of $f$ define the number density
\begin{equation}
\label{2.4} n({\bf r}, t)=\int \; d{\bf v}f({\bf r}, {\bf v},t),
\end{equation}
the flow velocity
\begin{equation}
\label{2.5} {\bf u}({\bf r}, t)=\frac{1}{n({\bf r}, t)}\int \;
d{\bf v} {\bf v} f({\bf r},{\bf v},t),
\end{equation}
and the {\em granular} temperature
\begin{equation}
\label{2.6} T({\bf r}, t)=\frac{m}{d n({\bf r}, t)}\int \; d{\bf
v} V^2({\bf r}, t) f({\bf r},{\bf v},t),
\end{equation}
where ${\bf V}({\bf r},t)\equiv {\bf v}-{\bf u}({\bf r}, t)$ is the peculiar velocity. The macroscopic balance
equations for density, momentum, and energy follow directly from Eq.\ ({\ref{2.1}) by multiplying with $1$,
$m{\bf v}$, and $\frac{1}{2}mv^2$ and integrating over ${\bf v}$:
\begin{equation}
\label{2.7} D_{t}n+n\nabla \cdot {\bf u}=0\;,
\end{equation}
\begin{equation}
\label{2.8} D_{t}u_i+(mn)^{-1}\nabla_j P_{ij}=0\;,
\end{equation}
\begin{equation}
\label{2.9} D_{t}T+\frac{2}{dn}\left(\nabla \cdot {\bf
q}+P_{ij}\nabla_j u_i\right) =-\zeta T\;,
\end{equation}
where $D_{t}=\partial _{t}+{\bf u}\cdot \nabla$. The microscopic
expressions for the pressure tensor ${\sf P}$, the heat flux ${\bf
q}$, and the cooling rate $\zeta$ are given, respectively, by
\begin{equation}
{\sf P}({\bf r}, t)=\int d{\bf v}\,m{\bf V}{\bf V}\,f({\bf r},{\bf
v},t),
 \label{2.10}
\end{equation}
\begin{equation}
{\bf q}({\bf r}, t)=\int d{\bf v}\,\frac{1}{2}m V^{2}{\bf V}\,
f({\bf r},{\bf v},t), \label{2.11}
\end{equation}
\begin{equation}
\label{2.12} \zeta({\bf r}, t)=-\frac{1}{dn({\bf r},t)T({\bf r},
t)}\int\, d{\bf v} mV^2J[{\bf r},{\bf v}|f(t)].
\end{equation}
The balance equations (\ref{2.7})--(\ref{2.9}) apply regardless of the details of the interaction model
considered. The influence of the collision model appears through the $\alpha$-dependence of the cooling rate and
of the momentum and heat fluxes. In particular, the cooling rate $\zeta$ is given by \cite{S03}
\begin{equation}
\label{2.12.1} \zeta=\frac{1-\alpha^2}{2d}\omega=\frac{d+2}{4d}(1-\alpha^2)\nu_0.
\end{equation}

Let us assume that the gas is under USF. This idealized macroscopic state is characterized by a constant
density, a uniform temperature, and a simple shear with the local velocity field given by
\begin{equation}
\label{2.15} u_i=a_{ij}r_j, \quad a_{ij}=a\delta_{ix}\delta_{jy},
\end{equation}
where $a$ is the {\em constant} shear rate. This linear velocity profile assumes no boundary layer near the
walls and is generated by the Lees--Edwards boundary conditions \cite{LE72}, which are simply periodic boundary
conditions in the local Lagrangian frame moving with the flow velocity \cite{DSBR86}. Since the heat flux is
zero in the USF problem, the balance equation for the energy (\ref{2.9}) reads
\begin{equation}
\label{2.15.1} \nu_0^{-1}\partial_t \ln T=-\zeta^*-\frac{2a^*}{d}P_{xy}^*,
\end{equation}
where $\zeta^*=\zeta/\nu_0$, $a^*=a/\nu_0$, $P_{xy}^*=P_{xy}/p$, $p=nT$ being the hydrostatic pressure. Equation
(\ref{2.15.1}) shows that the temperature changes in time due to the competition between two (opposite)
mechanisms: on the one hand, viscous (shear) heating and, on the other hand, energy dissipation in collisions.
The {\em reduced} shear rate $a^*$ is the nonequilibrium relevant parameter of the USF problem since it measures
the departure of the system from equilibrium. Note that, except for Model A ($q=0$), $a^*(t)\propto T(t)^{-q}$
is a function of time through its dependence on temperature. Since in the hydrodynamic regime $P_{xy}^*(t)$
depends on time only through its dependence on $a^*(t)$ \cite{GS03}, then for $q\neq 0$ a steady state is
eventually reached in the long time limit when both viscous heating and collisional cooling cancel each other
and the fluid autonomously seeks the temperature at which the above balance occurs. In this situation, $a^*$ and
$\alpha$ are not independent quantities but they are related through the steady state condition:
\begin{equation}
\label{2.16} a^* P_{xy}^*=-\frac{d}{2}\zeta^*.
\end{equation}
However, when $q=0$, the collision frequency $\nu_0$ is independent of temperature and $a^*$ remains constant in
time, so that there is no steady state (except if $a^*$ takes the specific value given by (\ref{2.16})).
Consequently, only in the case of Model A the reduced shear rate $a^*$ and the coefficient of restitution
$\alpha$ are in general {\em independent} parameters in the USF state. Note that, although the temperature
changes in time, the distribution of velocities relative to the thermal speed $\sqrt{2T/m}$ is expected to
reach, after a sufficient number of collisions per particle, a stationary form that only depends on the control
parameters $a^*$ and $\alpha$ \cite{AS07}. As a consequence, when the velocity moments are conveniently scaled
with the thermal speed, they reach stationary values after a kinetic transient regime \cite{SG06,SGD04}. These
steady values are nonlinear functions of both the (reduced) shear rate and the coefficient of restitution. More
details on the USF state for dissipative systems can be found in Ref.\ \cite{SG06}.

The USF problem is perhaps the nonequilibrium state most widely studied in the past few years both for granular
and conventional gases \cite{GS03,C90}. At a microscopic level, it becomes spatially homogeneous when the
velocities of the particles are referred to the Lagrangian frame of reference co-moving with the flow velocity
${\bf u}$ \cite{DSBR86}. In this frame, the one-particle distribution function adopts the {\em uniform} form,
$f({\bf r},{\bf v})\rightarrow f({\bf V})$, and the Boltzmann equation (\ref{2.1}) reads
\begin{equation}
\label{2.17} \left(\partial_t-aV_y\frac{\partial}{\partial V_x} \right)f({\bf V})=J\left[ {\bf V}|f,f\right] \;.
\end{equation}
Upon writing Eq.\ (\ref{2.17}) use has been made of the identity
\begin{equation}
\label{2.17.1} {\bf v}\cdot \nabla f=v_y\frac{\partial f}{\partial u_x}\frac{\partial u_x}{\partial
y}=-aV_y\frac{\partial f}{\partial V_x},
\end{equation}
where in the last step I have taken into account that $f$ depends on ${\bf u}$ through the peculiar velocity
${\bf V}$. The elements of the pressure tensor provide information on the rheological properties of the system.
These elements can be obtained by multiplying the Boltzmann equation (\ref{2.17}) by $m V_i V_j$ and integrating
over ${\bf V}$. The result is \cite{S03,SG06,GS07}
\begin{equation}
\label{2.18} \partial_t
P_{ij}+a_{i\ell}P_{j\ell}+a_{j\ell}P_{i\ell}=-\nu_{0|2}(P_{ij}-p\delta_{ij})-\zeta
p \delta_{ij},
\end{equation}
where
\begin{equation}
\label{2.19} \nu_{0|2}=\zeta+\frac{(1+\alpha)^2}{4}\nu_0.
\end{equation}

In the case of Model A ($q=0$), the set of first-order
differential equations \ (\ref{2.18}) can be exactly solved
\cite{SG06}. In terms of the reduced elements
$P_{ij}^*=P_{ij}(t)/p(t)$, the solution can be written as
\begin{equation}
\label{2.20} P_{xx}^*=\frac{1+2 d \gamma(\widetilde{a})}{1+2\gamma(\widetilde{a})}\;,\quad
P_{yy}^*=P_{zz}^*=\frac{1}{1+2\gamma(\widetilde{a})}\;,
\end{equation}
\begin{equation}
\label{2.21}
P_{xy}^*=-d\frac{\gamma(\widetilde{a})}{\widetilde{a}}=-\frac{\widetilde{a}}
{[1+2\gamma(\widetilde{a})]^2}\;,
\end{equation}
where
\begin{equation}
\label{2.21.1} \widetilde{a}=\frac{a}{\omega_{0|2}},\quad
\omega_{0|2}=\nu_{0|2}-\zeta=\frac{(1+\alpha)^2}{4}\nu_0,
\end{equation}
and $\gamma(\widetilde{a})$ is the real root of the cubic equation
\begin{equation}
\label{2.22} \gamma(1+2\gamma)^2=\frac{\widetilde{a}^2}{d},
\end{equation}
namely,
\begin{equation}
\label{2.23}
\gamma(\widetilde{a})=\frac{2}{3}\sinh^2\left[\frac{1}{6}\cosh
^{-1}\left(1+\frac{27}{d}\widetilde{a}^2\right)\right].
\end{equation}
Insertion of Eqs.\ (\ref{2.20}) and (\ref{2.21}) into Eq.\ (\ref{2.18}) yields
\begin{equation}
\label{2.23.1}
\partial_t \ln T=2\lambda,
\end{equation}
where I have called
\begin{equation}
\label{2.24} \lambda=2\gamma(\widetilde{a}) \omega_{0|2}-\zeta.
\end{equation}
In the elastic limit ($\alpha=1$), $\widetilde{a}=a^*$, and Eqs.\ (\ref{2.20})--(\ref{2.24}) reduce to the
well-known exact solution obtained long time ago by Ikenberry and Truesdell for Maxwell molecules \cite{IT56}.
Equation (\ref{2.23.1}) shows that $T(t)$ either grows or decays exponentially. The first situation happens if
$2\gamma(\widetilde{a}) \omega_{0|2}>\zeta$. In that case, the imposed shear rate is sufficiently large (or the
dissipation in collisions is sufficiently low) as to make the viscous heating effect dominate over the inelastic
cooling effect. The opposite happens if $2\gamma(\widetilde{a}) \omega_{0|2}<\zeta$.

Beyond rheological properties, the next nontrivial moments in the USF problem are the fourth-degree velocity
moments. These moments are needed to determine the transport properties of the gas in states close to USF, which
is the main goal of this paper. Very recently, the dependence of the fourth-degree moments on the shear rate and
the coefficient of restitution has been explicitly obtained for IMM \cite{SG06}. Their explicit expressions are
displayed in Appendix \ref{appA} for a three-dimensional gas ($d=3$). In a similar way to the case of elastic
Maxwell molecules \cite{GS03}, it has been shown that, for a given value of $\alpha$, the fourth-degree moments
are divergent for shear rates larger than a certain critical value $a_c(\alpha)$. This singular behavior of the
moments reflects the existence of an algebraic high-velocity tail in the distribution function. However, for
practical reasons, since in general the numerical value $a_c(\alpha)$ is rather large, nonlinear shearing
effects are still significant for $a<a_c$.

As said before, in the case of Model B ($q\neq 0$), after a transient regime the system achieves a steady state,
so that $\lambda=0$ and Eq.\ (\ref{2.24}) leads to
\begin{equation}
\label{2.24.1} \gamma(\widetilde{a})=\frac{d+2}{2d}\frac{1-\alpha}{1+\alpha}.
\end{equation}
The steady solution for the pressure tensor is still given by Eqs.\ (\ref{2.20}) and (\ref{2.21}), except that
$\widetilde{a}$ (or, equivalently $a^*$) and $\alpha$ are not independent quantities. The explicit dependence of
the steady-state value $a_s^*(\alpha)$ of the reduced shear rate can be easily obtained by inserting the
condition (\ref{2.24.1}) into Eq.\ (\ref{2.22}):
\begin{equation}
\label{2.25} a_s^*(\alpha)=\sqrt{\frac{d+2}{2}(1-\alpha^2)}\frac{d+1-\alpha}{2d}.
\end{equation}
It must be remarked that the steady-state results are ``universal'' in the sense that they hold both for Model A
and Model B, regardless of the precise dependence $\nu_0(t)$ \cite{SG06}. For elastic collisions, Eq.\
(\ref{2.25}) yields $a^*=0$ and so the equilibrium results are recovered, i.e., $P_{ij}^*=\delta_{ij}$. The
analytical results obtained for the steady USF state in the case of Model B \cite{G03} agree quite well with
Monte Carlo simulations of the Boltzmann equation for IHS \cite{BRM97,SA05}, even for strong dissipation.

\section{Small perturbations around the uniform shear flow state}
\label{sec3}

Let us assume now that we disturb the USF by small spatial perturbations. The response of the system to these
perturbations gives rise to additional contributions to the momentum and heat fluxes, which can be characterized
by generalized transport coefficients. Since the system is strongly sheared, the corresponding transport
coefficients are highly nonlinear functions of the shear rate. The goal here is to determine the shear-rate
dependence of these coefficients for IMM.

To analyze this problem, one has to start from the Boltzmann equation (\ref{2.1}) with a general time and space
dependence. First, it is convenient to keep using the relative velocity ${\bf V}={\bf v}-{\bf u}_0$, where ${\bf
u}_0={\sf a}\cdot {\bf r}$ is the flow velocity of the {\em undisturbed} USF state. Here, the only nonzero
element of the tensor ${\sf a}$ is $a_{ij}=a\delta_{ix}\delta_{jy}$. On the other hand, in the {\em disturbed}
state the true velocity ${\bf u}$ is in general different from ${\bf u}_0$, i.e., ${\bf u}={\bf u}_0+\delta {\bf
u}$, $\delta {\bf u}$ being a small perturbation to ${\bf u}_0$. As a consequence, the true peculiar velocity is
now  ${\bf c}\equiv {\bf v}-{\bf u}={\bf V}-\delta{\bf u}$. In the Lagrangian frame moving with velocity ${\bf
u}_0$, the convective operator ${\bf v}\cdot \nabla$ can be written as
\begin{equation}
\label{3.0} {\bf v}\cdot \nabla f=\left({\bf V}+{\bf u}_0\right) \cdot \nabla f=-aV_y\frac{\partial}{\partial
V_x}f+\left({\bf V}+{\bf u}_0\right) \cdot \nabla f,
\end{equation}
where the derivative $\nabla f$ in the last term must be taken now at constant ${\bf V}$. According to the
identity (\ref{3.0}), the Boltzmann equation reads
\begin{equation}
\label{3.1} \frac{\partial}{\partial t}f-aV_y\frac{\partial}{\partial V_x}f+\left({\bf V}+{\bf u}_0\right) \cdot
\nabla f=J[{\bf V}|f,f].
\end{equation}
I am interested in computing the transport coefficients in a state that slightly deviates from the USF. For this
reason, I assume that the spatial gradients of the hydrodynamic fields
\begin{equation}
\label{3.1.1} A({\bf r},t)\equiv \{n({\bf r},t), T({\bf r}, t),
\delta {\bf u}({\bf r},t)\}
\end{equation}
are small. Under these conditions, a solution to the Boltzmann equation (\ref{3.1}) can be obtained by means of
a generalization of the conventional Chapman--Enskog method \cite{CC70}, where the velocity distribution
function is expanded about a {\em local} shear flow reference state in terms of the small spatial gradients of
the hydrodynamic fields relative to those of USF. This type of Chapman--Enskog-like expansion has been
considered in the case of elastic gases to get the set of shear-rate dependent transport coefficients
\cite{GS03,LD97} in a thermostatted shear flow problem and it has also been recently considered
\cite{G06,L06,G07} in the context of inelastic gases.

The Chapman--Enskog method assumes the existence of a {\em normal} solution in which all space and time
dependence of the distribution function occurs through a functional dependence on the fields $A({\bf r},t)$,
i.e.,
\begin{equation}
\label{3.2} f({\bf r}, {\bf V},t)\equiv f[{\bf V}|A({\bf r}, t)].
\end{equation}
This solution expresses the fact that the space dependence of the
shear flow is completely absorbed in the relative velocity ${\bf
V}$ and all other space and time dependence occurs entirely
through a functional dependence on the fields ${\bf A}({\bf
r},t)$. This functional dependence can be made local by an
expansion of the distribution function in powers of the
hydrodynamic gradients:
\begin{equation}
\label{3.3} f({\bf r}, {\bf V},t) =f^{(0)}({\bf V}|A({\bf r}, t))+ f^{(1)}({\bf V}|A({\bf r}, t))+\cdots,
\end{equation}
where the reference zeroth-order distribution function corresponds to the USF distribution function but taking
into account the local dependence of the density and temperature and the change ${\bf V}\rightarrow {\bf c}={\bf
V}-\delta{\bf u}({\bf r}, t)$. The new feature of this Chapman--Enskog expansion (in contrast to the
conventional one) is that the successive approximations $f^{(k)}$ are of order $k$ in the gradients of $n$, $T$,
and $\delta {\bf u}$ but retain all the orders in the shear rate $a$. More technical details on this
Chapman--Enskog-like type of expansion can be found in Appendix \ref{appB}.

The expansion (\ref{3.3}) yields the corresponding expansion for
the fluxes and the cooling rate when one substitutes (\ref{3.3})
into their definitions (\ref{2.10})--(\ref{2.12}):
\begin{equation}
\label{3.4} {\sf P}={\sf P}^{(0)}+{\sf P}^{(1)}+\cdots, \quad {\bf
q}={\bf q}^{(0)}+{\bf q}^{(1)}+\cdots, \quad
\zeta=\zeta^{(0)}+\zeta^{(1)}+\cdots.
\end{equation}
Finally, as in the usual Chapman--Enskog method, the time derivative is also expanded as
\begin{equation}
\label{3.5}
\partial_t=\partial_t^{(0)}+\partial_t^{(1)}+\partial_t^{(2)}+\cdots,
\end{equation}
where the action of each operator $\partial_t^{(k)}$ is also
defined in Appendix \ref{appB}. In this paper, only the zeroth and
first order approximations will be considered.

\subsection{Zeroth-order approximation}

Substituting the expansions (\ref{3.3}) and (\ref{3.5}) into Eq.\
(\ref{3.1}), the kinetic equation for $f^{(0)}$ is given by
\begin{equation}
\label{3.6}
\partial_t^{(0)}f^{(0)}-aV_y\frac{\partial}{\partial V_x}f^{(0)}=J[{\bf V}|f^{(0)},f^{(0}].
\end{equation}
To lowest order in the expansion the conservation laws give
\begin{equation}
\label{3.7}
\partial_t^{(0)}n=0,\quad \partial_t^{(0)}T=-\frac{2}{dn}a P_{xy}^{(0)}-T\zeta,
\end{equation}
\begin{equation}
\label{3.8}
\partial_t^{(0)}\delta u_i+a_{ij} \delta u_j=0.
\end{equation}
Upon writing the second identity in Eq.\ (\ref{3.7}) I have taken into account that the effective collision
frequency $\omega \propto n T^q$ is assumed to be a functional of $f$ only through the density and temperature.
Consequently, $\omega^{(0)}=\omega$, $\omega^{(1)}=\omega^{(2)}=\cdots=0$ and, using Eq.\ (\ref{2.12.1}),
$\zeta^{(0)}=\zeta$, $\zeta^{(1)}=\zeta^{(2)}=\cdots=0$. It must be noticed that, in the case of IHS,
$\zeta^{(1)}$ is different from zero but quite small \cite{G06}.

Since $f^{(0)}$ is a normal solution, its dependence on time only
occurs through $n$, $\delta {\bf u}$ and $T$:
\begin{eqnarray}
\label{3.9}
\partial_t^{(0)}f^{(0)}&=&\frac{\partial f^{(0)}}{\partial
n}\partial_t^{(0)} n+\frac{\partial f^{(0)}}{\partial
T}\partial_t^{(0)} T+\frac{\partial f^{(0)}}{\partial \delta
u_i}\partial_t^{(0)} \delta u_i\nonumber\\
&=&-\left(\frac{2}{d n}a
P_{xy}^{(0)}+T\zeta\right)\frac{\partial}{\partial
T}f^{(0)}-a_{ij}\delta u_j
\frac{\partial}{\partial \delta u_i}f^{(0)}\nonumber\\
&=&-\left(\frac{2}{d n}a
P_{xy}^{(0)}+T\zeta\right)\frac{\partial}{\partial
T}f^{(0)}+a_{ij}\delta u_j \frac{\partial}{\partial c_i}f^{(0)},
\end{eqnarray}
where I have taken into account again that $f^{^(0)}$ depends on $\delta {\bf u}$ through ${\bf c}$.
Substitution of Eq.\ (\ref{3.9}) into Eq.\ (\ref{3.6}) yields the following kinetic equation for $f^{(0)}$:
\begin{equation}
\label{3.10} -\left(\frac{2}{d n}a
P_{xy}^{(0)}+T\zeta\right)\frac{\partial}{\partial T}f^{(0)}
-ac_y\frac{\partial}{\partial c_x}f^{(0)}=J[{\bf
V}|f^{(0)},f^{(0}].
\end{equation}
The zeroth-order solution leads to ${\bf q}^{(0)}={\bf 0}$. In the case of Model A, the zeroth-order pressure
tensor is given by Eqs.\ (\ref{2.20}) and (\ref{2.21}) while the expressions of the fourth-degree moments are
displayed in Appendix \ref{appA}. As shown in Refs.\ \cite{G06,L06,G07}, for given values of $a$ and $\alpha$,
the steady state condition (\ref{2.16}) establishes a mapping between the density and temperature so that every
density corresponds to one and only one temperature. Since the density $n({\bf r}, t)$ and temperature $T({\bf
r}, t)$ are specified separately in the {\em local} USF state, the viscous heating only partially compensates
for the collisional cooling and so, $\partial_t^{(0)} T \neq 0$ \cite{L06}. Consequently, the zeroth-order
distribution $f^{(0)}$ depends on time through its dependence on temperature and the quantities $a^*$ and
$\alpha$ must be considered as independent parameters for general infinitesimal perturbations around the USF
state. This fact gives rise to new and conceptual practical difficulties not present in the previous analysis
made for elastic thermostatted gases \cite{LD97}.

The set of equations for $P_{ij}^{(0)}$ follows from Eq.\
(\ref{3.10}) as
\begin{equation}
\label{3.11} -\left(\frac{2}{d n}a
P_{xy}^{(0)}+T\zeta\right)\frac{\partial}{\partial T}
P_{ij}^{(0)}+a_{i\ell}P_{j\ell}^{(0)}+a_{j\ell}P_{i\ell}^{(0)}=-\nu_{0|2}(P_{ij}^{(0)}-
p\delta_{ij})-\zeta p \delta_{ij}.
\end{equation}
The dependence of $P_{ij}^{(0)}$ on temperature occurs explicitly
through the hydrostatic pressure $p=nT$ and through its dependence
on $a^*$. Consequently,
\begin{equation}
\label{3.12} T\frac{\partial}{\partial T} P_{ij}^{(0)}=
T\frac{\partial}{\partial T}p P_{ij}^*(a^*)=p\left(1-q
a^*\frac{\partial}{\partial a^*}\right)P_{ij}^*(a^*),
\end{equation}
where here $P_{ij}^*=P_{ij}^{(0)}/p$. Thus, in dimensionless form Eq.\ (\ref{3.11}) becomes
\begin{equation}
\label{3.13} -\left(\frac{2}{d}a^* P_{xy}^{*}+\zeta^*\right)\left(1-q a^*\frac{\partial}{\partial
a^*}\right)P_{ij}^* +a_{i\ell}^*P_{j\ell}^{*}+a_{j\ell}^*P_{i\ell}^{*}=-\nu_{0|2}^*(P_{ij}^{*}-
\delta_{ij})-\zeta^* \delta_{ij},
\end{equation}
where $\nu_{0|2}^*=\nu_{0|2}/\nu_0$.  For $q=0$ (Model A), the solution to Eq.\ (\ref{3.13}) is given by Eqs.\
(\ref{2.20}) and (\ref{2.21}). For $q\neq 0$ (Model B), Eq.\ (\ref{3.13}) must be solved numerically to get the
dependence of $P_{ij}^*$ on $a^*$. In the case of IHS ($q=\frac{1}{2}$), a detailed study of the dependence of
$P_{ij}^*$ on $a^*$ has been carried out in Ref.\ \cite{SGD04}. For elastic gases ($\alpha=1$), a comparison
between the results derived for Maxwell molecules ($q=0$) and hard spheres ($q=\frac{1}{2}$) at the level of
rheological properties has shown that both results are very close \cite{GS03} so that there is a weak influence
of the interaction model on transport properties. This suggests to expand the transport properties in powers of
the interaction parameter $q$ as an alternative to get accurate analytical results for non-Maxwell molecules.
For inelastic collisions, I expect that such a good agreement is also kept and so the following formal expansion
in $q$ is considered:
\begin{equation}
\label{3.14} P_{ij}^*(a^*)=P_{ij,A}^*(a^*)+q \Delta P_{ij}^*(a^*)+\cdots,
\end{equation}
where $P_{ij,A}^*(a^*)$ is the known result for Model A. Inserting
Eq.\ (\ref{3.14}) into Eq.\ (\ref{3.13}) and neglecting terms
nonlinear in $q$, one gets
\begin{equation}
\label{3.15} \Delta P_{yy}^*=-\frac{6\lambda^*}{\omega_{0|2}^*}\frac{\gamma}{(1+2\gamma)(1+6\gamma)^2},
\end{equation}
\begin{equation}
\label{3.16} \Delta
P_{xy}^*=-\frac{\lambda^*\widetilde{a}}{\omega_{0|2}^*}\frac{1-6\gamma}{(1+2\gamma)^2(1+6\gamma)^2},
\end{equation}
where $\omega_{0|2}^*=\omega_{0|2}/\nu_0$ and $\lambda^*=\lambda/\nu_0$, $\lambda$ being given by Eq.\
(\ref{2.24}). Upon writing Eqs.\ (\ref{3.15}) and (\ref{3.16}) use has been made of the relations
\begin{equation}
\label{3.16.1} a^*\frac{\partial}{\partial
a^*}P_{yy,A}^*=-\frac{4\gamma}{(1+2\gamma)(1+6\gamma)},
\end{equation}
\begin{equation}
\label{3.16.2} a^*\frac{\partial}{\partial
a^*}P_{xy,A}^*=-\widetilde{a}\frac{1-2\gamma}{(1+2\gamma)^2(1+6\gamma)}.
\end{equation}
\begin{figure}
\includegraphics[width=0.6 \columnwidth,angle=0]{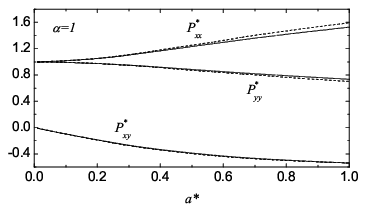}
\caption{Shear-rate dependence of the elements of the reduced
pressure tensor $P_{ij}^*=P_{ij}^{(0)}/p$ for $\alpha=1$ in the
three-dimensional case. The solid lines correspond to the results
obtained for Model A ($q=0$) while the dashed lines refer to the
results derived for Model B for $q=\frac{1}{2}$ by using the
approximation (\ref{3.14}). \label{fig01}}
\end{figure}
\begin{figure}
\includegraphics[width=0.6 \columnwidth,angle=0]{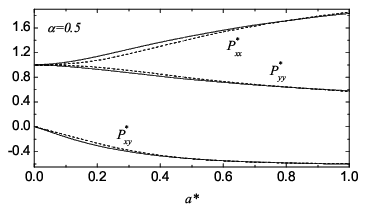}
\caption{Shear-rate dependence of the elements of the reduced
pressure tensor $P_{ij}^*=P_{ij}^{(0)}/p$ for $\alpha=0.5$ in the
three-dimensional case. The solid lines correspond to the results
obtained for Model A ($q=0$) while the dashed lines refer to the
results derived for Model B for $q=\frac{1}{2}$ by using the
approximation (\ref{3.14}). \label{fig02}}
\end{figure}
From Eq.\ (\ref{3.14}) and taking into account Eqs.\ (\ref{2.20})
and (\ref{2.21}), the expressions for the elements of the pressure
tensor for Model B can be written as
\begin{equation}
\label{3.16.3} P_{xx}^*=\frac{1}{1+2\gamma}\left[1+2d\gamma-q\frac{6\gamma\lambda^*}{\omega_{0|2}^*}
\frac{1-d}{(1+6\gamma)^2}\right],
\end{equation}
\begin{equation}
\label{3.16.5}
P_{yy}^*=P_{zz}^*=\frac{1}{1+2\gamma}\left[1-q\frac{6\gamma\lambda^*}{\omega_{0|2}^*(1+6\gamma)^2}\right],
\end{equation}
\begin{equation}
\label{3.16.4}
P_{xy}^*=-\frac{\widetilde{a}}{(1+2\gamma)^2}\left[1+q\frac{\lambda^*}{\omega_{0|2}^*}
\frac{1-6\gamma}{(1+6\gamma)^2}\right].
\end{equation}
\begin{figure}
\includegraphics[width=0.6 \columnwidth,angle=0]{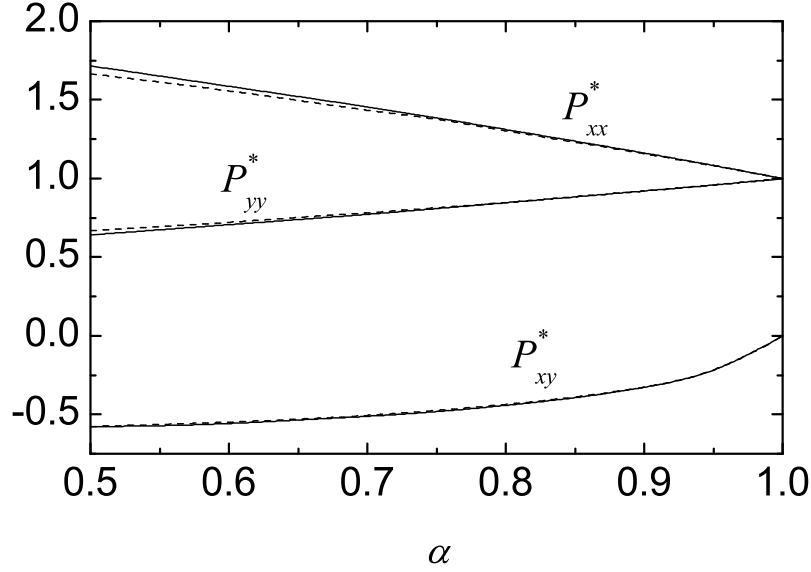}
\caption{Plot of the reduced elements of the pressure tensor ${\sf P}^*$ as functions of the coefficient of
restitution $\alpha$ in the steady USF state. The solid lines are the results derived here for IMM while the
dashed lines correspond to the results obtained for IHS in Ref.\ \cite{BRM97}. \label{fig1}}
\end{figure}
As expected, when the steady state condition (\ref{2.14}) applies locally, then $\partial_t^{(0)}T=0$,
$\lambda^*=0$ so that, according to Eqs.\ (\ref{3.15}) and (\ref{3.16}), the results for the pressure tensor are
independent of the model interaction considered. Note that the same type of approximation (\ref{3.14}) can be
used to estimate the fourth-degree moments for $q\neq 0$, although this calculation will be omitted here for
simplicity.

Figures \ref{fig01} and \ref{fig02} show the dependence of the reduced pressure tensor $P_{ij}^*$ on the reduced
shear rate $a^*$ for a three-dimensional system ($d=3$) and two different values of the coefficient of
restitution: $\alpha=1$ and $\alpha=0.5$. I have considered the exact results obtained for Model A and the
results for Model B (when $q=\frac{1}{2}$) by using the approximation (\ref{3.14}). It is quite apparent that,
for a given interaction model, there is a weak influence of the inelasticity on the shear-rate dependence of the
pressure tensor elements. Furthermore, for a given value of $\alpha$, the differences between the results
obtained for Model A and Model B are very small in the range of shear rates considered. This means that the
rheological properties are quite insensitive to the interaction model considered, even for strong dissipation.
The dependence of the pressure tensor on dissipation in the steady state [i.e., when the (reduced) shear rate
and the coefficient of restitution $\alpha$ are not independent parameters but they are related through Eq.\
(\ref{2.25})] is plotted in Fig.\ \ref{fig1} along with the results obtained for IHS \cite{BRM97,SGD04}. It can
be observed that the IMM results reproduce very well the IHS predictions over the range of values of the
coefficient of restitution analyzed.

\subsection{First-order approximation}

The analysis to first order in the gradients is worked out in
Appendix \ref{appB}. Only the final results are presented in this
Section. The distribution function $f^{(1)}$ is of the form
\begin{equation}
\label{3.17} f^{(1)}={\bf X}_{n}\cdot \nabla n+ {\bf X}_{T}\cdot
\nabla T+{\sf X}_{u}:\nabla \delta {\bf u},
\end{equation}
where the vectors ${\bf X}_{n}$ and ${\bf X}_{T}$ and the tensor
${\sf X}_{u}$ are functions of the true peculiar velocity ${\bf
c}$. They are the solutions of the following linear integral
equations:
\begin{equation}
\label{3.18} -\left[\left(\frac{2}{d n}a
P_{xy}^{(0)}+T\zeta\right)\partial_T+ a
c_y\frac{\partial}{\partial c_x}-{\cal
L}\right]X_{n,i}+\frac{T}{n}\left[\frac{2a}{dp}(1-n\partial_n)
P_{xy}^{(0)}-\zeta\right]X_{T,i}=Y_{n,i},
\end{equation}
\begin{equation}
\label{3.19} -\left[\left(\frac{2}{dn}a
P_{xy}^{(0)}+T\zeta\right)\partial_T+ \frac{2a}{dp}T(\partial_T
P_{xy}^{(0)})+(q+1)\zeta+a c_y\frac{\partial}{\partial c_x}-{\cal
L}\right]X_{T,i}=Y_{T,i},
\end{equation}
\begin{equation}
\label{3.20} -\left[\left(\frac{2}{dn}a
P_{xy}^{(0)}+T\zeta\right)\partial_T+a c_y\frac{\partial}{\partial
c_x}-{\cal L}\right]X_{u,k\ell}-
a\delta_{ky}X_{u,x\ell}=Y_{u,k\ell},
\end{equation}
where ${\bf Y}_n({\bf c})$, ${\bf Y}_T({\bf c})$, and ${\sf Y}_u({\bf c})$ are defined by Eqs.\
(\ref{b10})--(\ref{b12}), respectively. In addition, ${\cal L}$ is the linearized Boltzmann collision operator
around the reference USF state:
\begin{equation}
\label{3.21} {\cal L}X\equiv
-\left(J[f^{(0)},X]+J[X,f^{(0)}]\right).
\end{equation}
It is worth noting that for $q=\frac{1}{2}$, Eqs.\
(\ref{3.18})--(\ref{3.20}) have the same structure as that of the
Boltzmann equation for IHS \cite{G06}. The only difference between
both models lies in the explicit form of the linearized operator
${\cal L}$.

With the distribution function $f^{(1)}$ determined by Eq.\
(\ref{3.17}), the first-order corrections to the fluxes are
\begin{equation}
\label{3.22} P_{ij}^{(1)}=-\eta_{ijk\ell} \frac{\partial \delta
u_k} {\partial r_{\ell}},
\end{equation}
\begin{equation}
\label{3.23} q_i^{(1)}=-\kappa_{ij}\frac{\partial T}{\partial
r_j}- \mu_{ij}\frac{\partial n}{\partial r_j},
\end{equation}
where
\begin{equation}
\label{3.24} \eta_{ijk\ell}=-\int\; d{\bf c}\, mc_ic_j
X_{u,k\ell}({\bf c}),
\end{equation}
\begin{equation}
\label{3.25} \kappa_{ij}=-\int\; d{\bf c}\, \frac{m}{2}c^2c_i
X_{T,j}({\bf c}),
\end{equation}
\begin{equation}
\label{3.26} \mu_{ij}=-\int\; d{\bf c}\, \frac{m}{2}c^2c_i
X_{n,j}({\bf c}).
\end{equation}
Upon writing Eqs.\ (\ref{3.22})--(\ref{3.26}) use has been made of
the symmetry properties of $X_{n,i}$, $X_{T,i}$ and $X_{u,ij}$. In
general, the set of {\em generalized} transport coefficients
$\eta_{ijk\ell}$, $\kappa_{ij}$, and $\mu_{ij}$ are nonlinear
functions of the coefficient of restitution $\alpha$ and the
reduced shear rate $a^*$. The anisotropy induced in the system by
the presence of shear flow gives rise to new transport
coefficients, reflecting broken symmetry. The momentum flux is
expressed in terms of a viscosity tensor $\eta_{ijk\ell}(a^*,
\alpha)$ of rank 4 which is symmetric and traceless in $ij$ due to
the properties of the pressure tensor $P_{ij}^{(1)}$. The heat
flux is expressed in terms of a thermal conductivity tensor
$\kappa_{ij}(a^*, \alpha)$ and a new tensor $\mu_{ij}(a^*,
\alpha)$. Of course, for $a^*=0$ and $\alpha=1$, the usual
Navier-Stokes constitutive equations for ordinary gases are
recovered and the transport coefficients become
\begin{equation}
\label{3.27} \eta_{ijk\ell}\rightarrow
\eta_0\left(\delta_{ik}\delta_{j\ell}+\delta_{jk}\delta_{i\ell}-
\frac{2}{d}\delta_{ij}\delta_{k\ell}\right),\quad
\kappa_{ij}\rightarrow \kappa_0 \delta_{ij}, \quad
\mu_{ij}\rightarrow 0,
\end{equation}
where $\eta_0=p/\nu_0$ and $\kappa_0=d(d+2) \eta_0/2(d-1)m$ are
the shear viscosity and thermal conductivity coefficients for
elastic collisions \cite{CC70}.

\section{Shear-rate dependent transport coefficients}
\label{sec4}

This Section is devoted to the determination of the generalized
transport coefficients $\eta_{ijk\ell}$, $\kappa_{ij}$, and
$\mu_{ij}$ associated with the momentum and heat fluxes. Let us
consider each flux separately.

\subsection{Momentum flux}

To first order in the hydrodynamic gradients, the momentum flux is given by Eq.\ (\ref{3.22}}). To get the
coefficient $\eta_{ijk\ell}$, I multiply both sides of Eq.\ (\ref{3.20}) by $mc_ic_j$ and integrate over ${\bf
c}$. The result is
\begin{eqnarray}
\label{4.1} \left(\frac{2}{d n}a P_{xy}^{(0)}+T\zeta\right)
\partial_T\eta_{ijk\ell}&-&a\left(\delta_{ix}\eta_{jyk\ell}+
\delta_{jx}\eta_{iyk\ell}-\delta_{ky}\eta_{ijx\ell}\right)
-\nu_{0|2}\eta_{ijk\ell}\nonumber\\
&=&-\delta_{k\ell}\left(1-n\partial_n\right)P_{ij}^{(0)}-
\left(\delta_{ik}P_{j\ell}^{(0)}+\delta_{jk}P_{i\ell}^{(0)}\right)
\nonumber\\
& &  +\frac{2}{d
n}\left(P_{k\ell}^{(0)}-a\eta_{xyk\ell}\right)\partial_TP_{ij}^{(0)}.
\end{eqnarray}
Upon writing Eq.\ (\ref{4.1}), use has been made of the results \cite{S03,GS07}
\begin{equation}
\label{4.2} \int \; d{\bf c}\; mc_ic_j{\cal
L}X_{u,k\ell}=-\nu_{0|2}\eta_{ijk\ell},
\end{equation}
\begin{eqnarray}
\label{4.3} \int \; d{\bf c}\;
mc_ic_jY_{u,k\ell}&=&-\delta_{k\ell}\left(1-n\partial_n\right)P_{ij}^{(0)}-
\left(\delta_{ik}P_{j\ell}^{(0)}+\delta_{jk}P_{i\ell}^{(0)}\right)\nonumber\\
& &  +\frac{2}{d
n}\left(P_{k\ell}^{(0)}-a\eta_{xyk\ell}\right)\partial_TP_{ij}^{(0)}.
\end{eqnarray}
The generalized shear viscosity can be written as $\eta_{ijk\ell}=\eta_0 \eta_{ijk\ell}^*(a^*)$ where
$\eta_{ijk\ell}^*(a^*)$ is a dimensionless function of the reduced shear rate $a^*$ and the coefficient of
restitution $\alpha$. The dependence of $\eta_{ijk\ell}^*$ on temperature is through the reduced shear rate
$a^*$ and so
\begin{equation}
\label{4.4} T\partial_T\eta_{ijk\ell}=T\partial_T \eta_0
\eta_{ijk\ell}^*(a^*)=(1-q)\eta_{ijk\ell}-q\eta_{ijk\ell}
a^*\partial_{a^*}\ln \eta_{ijk\ell}^*(a^*).
\end{equation}
Consequently, in dimensionless form, Eq.\ (\ref{4.1}) yields
\begin{eqnarray}
\label{4.5} \left(\frac{2}{d}a^* P_{xy}^{*}+\zeta^*\right)
\left[1-q(1+a^*\partial_{a^*})\right]\eta_{ijk\ell}^*
&-&a^*\left(\delta_{ix}\eta_{jyk\ell}^*+
\delta_{jx}\eta_{iyk\ell}^*-\delta_{ky}\eta_{ijx\ell}^*\right)
-\nu_{0|2}^*\eta_{ijk\ell}^*\nonumber\\
&=&-\delta_{k\ell}a^*\partial_{a^*}P_{ij}^{*}-
\left(\delta_{ik}P_{j\ell}^{*}+\delta_{jk}P_{i\ell}^{*}\right)\nonumber\\
& &
+\frac{2}{d}\left(P_{k\ell}^{*}-a^*\eta_{xyk\ell}^*\right)\left(1-qa^*\partial_{a^*}\right)
P_{ij}^{*},
\end{eqnarray}
where use has been made of the identity
\begin{equation}
\label{4.6} n\frac{\partial}{\partial n} P_{ij}^{(0)}=
n\frac{\partial}{\partial n}p P_{ij}^*(a^*)=p\left(1-
a^*\frac{\partial}{\partial a^*}\right)P_{ij}^*(a^*).
\end{equation}

In the absence of shear field ($a^*=0$), $P_{ij}^*=\delta_{ij}$,
and so Eq.\ (\ref{4.5}) has the solution
\begin{equation}
\label{4.6.1}
\eta_{ijk\ell}^*=\left[\nu_{0|2}^*-\zeta^*(1-q)\right]^{-1}
\left(\delta_{ik}\delta_{j\ell}+\delta_{jk}\delta_{i\ell}-
\frac{2}{d}\delta_{ij}\delta_{k\ell}\right).
\end{equation}
This expression coincides with the one derived by Santos \cite{S03} for IMM for vanishing shear rates. Beyond
this limit case, in general Eq.\ (\ref{4.5}) is a nonlinear differential equation that must be solved with the
appropriate boundary conditions to get the hydrodynamic solution. The simplest model is that of Model A, in
which case the set (\ref{4.5}) becomes a set of coupled algebraic equations.

\subsubsection{Model A}

When $q=0$, then the shear viscosity obeys the equation
\begin{eqnarray}
\label{4.7} \left(\frac{2}{d}a^* P_{xy}^{*}-\omega_{0|2}^*\right) \eta_{ijk\ell}^*
&-&a^*\left(\delta_{ix}\eta_{jyk\ell}^*+ \delta_{jx}\eta_{iyk\ell}^*-\delta_{ky}\eta_{ijx\ell}^*\right)
\nonumber\\
&=&-\delta_{k\ell}a^*\partial_{a^*}P_{ij}^{*}-
\left(\delta_{ik}P_{j\ell}^{*}+\delta_{jk}P_{i\ell}^{*}\right)\nonumber\\
& & +\frac{2}{d}\left(P_{k\ell}^{*}-a^*\eta_{xyk\ell}^*\right) P_{ij}^{*}.
\end{eqnarray}
The explicit dependence of $\eta_{ijk\ell}^*$ on $a^*$ and $\alpha$ can be obtained by solving the set of
algebraic equations (\ref{4.7}). As an example, the coefficients of the form $\eta_{ijxy}^*$ are obtained in
detail in Appendix \ref{appC}. The nonzero elements of $\eta_{ijxy}^*$ are given by
\begin{equation}
\label{4.8}
\eta_{xyxy}^*=\frac{1}{\omega_{0|2}^*}\frac{1-2\gamma}{(1+2\gamma)^2(1+6\gamma)},
\end{equation}
\begin{equation}
\label{4.8.1} \eta_{xxxy}^*=\frac{4}{\omega_{0|2}^*}\frac{1-d}{d}
\frac{\widetilde{a}}{(1+2\gamma)^3(1+6\gamma)},
\end{equation}
\begin{equation}
\label{4.9} \eta_{yyxy}^*=\eta_{zzxy}^*=\frac{4}{d\omega_{0|2}^*}
\frac{\widetilde{a}}{(1+2\gamma)^3(1+6\gamma)}.
\end{equation} When $\alpha=1$, Eq.\ (\ref{4.8}) reduces to the one previously derived for elastic gases
\cite{LD97}.

\subsubsection{Model B}

When $q\neq 0$, the coefficients $\eta_{ijk\ell}^*$ must be
obtained numerically. However, assuming that the dependence of
$\eta_{ijk\ell}^*$ on $a^*$ and $\alpha$ for $q\neq 0$ does not
differ significantly from the one obtained for Model A ($q=0$),
one can consider again the approximation
\begin{equation}
\label{4.10} \eta_{ijk\ell}^*= \eta_{ijk\ell,A}^*+q\Delta \eta_{ijk\ell}^*+\cdots,
\end{equation}
to get analytical results for $\eta_{ijk\ell}^*$. In Eq.\ (\ref{4.10}), $\eta_{ijk\ell,A}^*$ refers to the known
result for Model A. The unknown $\Delta \eta_{ijk\ell}^*$ can be determined by inserting (\ref{4.10}) into Eq.\
(\ref{4.5}) and retaining only linear terms in $q$. As an illustration, the quantities $\Delta \eta_{ijxy}^*$
are also evaluated in Appendix \ref{appC}. Their explicit expressions are given by Eqs.\
(\ref{c4.1})--(\ref{c4.4}).

\subsection{Heat Flux}

The heat flux is defined by Eq.\ (\ref{3.23}) in terms of the
coefficients $\kappa_{ij}$, Eq.\ (\ref{3.25}), and $\mu_{ij}$,
Eq.\ (\ref{3.26}). To get these generalized Navier-Stokes
coefficients, let us introduce the tensors
\begin{equation}
\label{4.11} \kappa_{ijk\ell}=-\int\; d{\bf c}\, \frac{m}{2}c_i
c_j c_k X_{T,\ell}({\bf c}),
\end{equation}
\begin{equation}
\label{4.12} \mu_{ijk\ell}=-\int\; d{\bf c}\, \frac{m}{2}c_i c_j
c_k  X_{n,\ell}({\bf c}).
\end{equation}
The generalized thermal conductivity $\kappa_{ij}$ and the new coefficient $\mu_{ij}$ are given by
\begin{equation}
\label{4.13} \kappa_{ij}=\kappa_{kkij},\quad \mu_{ij}=\mu_{kkij}.
\end{equation}
\begin{figure}
\includegraphics[width=0.6 \columnwidth,angle=0]{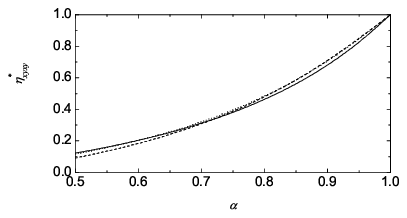}
\caption{Plot of the reduced element $\eta_{xyxy}^*$ as function of the coefficient of restitution $\alpha$ in
the steady USF state for a three-dimensional system. The solid line is the result derived here for  Model A
($q=0$), the dashed line refers to the result for Model B for $q=\frac{1}{2}$ by using the approximation
(\ref{4.10}), and the dotted line corresponds to the result obtained for IHS in Ref.\ \cite{G06}. \label{fig2}}
\end{figure}

Let us consider first the tensor $\kappa_{ijk\ell}$. To get it, I multiply both sides of Eq.\ (\ref{3.19}) by
$\frac{m}{2}c_ic_jc_k$ and integrate over velocity. After some algebra, one gets
\begin{eqnarray}
\label{4.14} \left(\frac{2}{d n}a P_{xy}^{(0)}+T\zeta\right)
\partial_T\kappa_{ijk\ell}&+&\left(\frac{2a}{d
p}(T\partial_TP_{xy}^{(0)})+(q+1)\zeta-\nu_{0|3}\right)\kappa_{ijk\ell}
\nonumber\\
& & - a\left(\delta_{ix}\kappa_{jky\ell}+
\delta_{jx}\kappa_{iky\ell}+\delta_{kx}\kappa_{ijy\ell}\right)\nonumber\\
& &
-\frac{\nu_{2|1}-\nu_{0|3}}{d+2}\left(\delta_{jk}\kappa_{i\ell}+
\delta_{ik}\kappa_{j\ell}+\delta_{ij}\kappa_{k\ell}\right)=
-\frac{m}{2}\partial_TN_{ijk\ell}^{(0)}\nonumber\\
&+&\frac{1}{2m n}
\left(P_{kj}^{(0)}\partial_TP_{i\ell}^{(0)}+P_{ik}^{(0)}\partial_TP_{j\ell}^{(0)}+
P_{ij}^{(0)}\partial_TP_{k\ell}^{(0)}\right),
\end{eqnarray}
where \cite{GS07}
\begin{equation}
\label{4.15} \nu_{2|1}=\frac{(1+\alpha)[5d+4-(d+8)\alpha]}{8d}\nu_0,\quad \nu_{0|3}=\frac{3}{2}\nu_{0|2},
\end{equation}
\begin{equation}
\label{4.15.1} N_{ijk\ell}^{(0)}=\int d{\bf c}\, c_i c_j c_k
c_{\ell}f^{(0)}({\bf c}).
\end{equation}
Upon writing Eq.\ (\ref{4.14}), use has been made of the results
\begin{equation}
\label{4.16} \int \; d{\bf c}\; \frac{m}{2}c_ic_jc_k{\cal
L}X_{T,\ell}=-\nu_{0|3}\kappa_{ijk\ell}-\frac{\nu_{2|1}-\nu_{0|3}}{d+2}\left(\delta_{jk}\kappa_{i\ell}+
\delta_{ik}\kappa_{j\ell}+\delta_{ij}\kappa_{k\ell}\right),
\end{equation}
\begin{eqnarray}
\label{4.17} \int \; d{\bf c}\;
\frac{m}{2}c_ic_jc_kY_{T,\ell}&=&-\frac{m}{2}\partial_TN_{ijk\ell}^{(0)}\nonumber\\
& & +\frac{1}{2m n}
\left(P_{kj}^{(0)}\partial_TP_{i\ell}^{(0)}+P_{ik}^{(0)}\partial_TP_{j\ell}^{(0)}+
P_{ij}^{(0)}\partial_TP_{k\ell}^{(0)}\right).
\end{eqnarray}
\begin{figure}
\includegraphics[width=0.6 \columnwidth,angle=0]{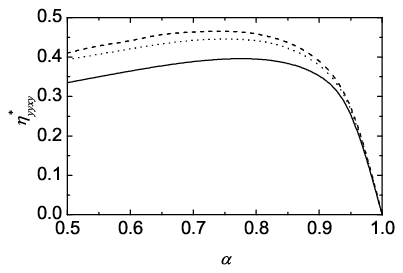}
\caption{Plot of the reduced element $\eta_{yyxy}^*$ as function of the coefficient of restitution $\alpha$ in
the steady USF state for a three-dimensional system. The solid line is the result derived here for  Model A
($q=0$), the dashed line refers to the result for Model B for $q=\frac{1}{2}$ by using the approximation
(\ref{4.10}), and the dotted line corresponds to the results obtained for IHS in Ref.\ \cite{G06}. \label{fig3}}
\end{figure}

As happens in the case of the shear viscosity, the thermal conductivity tensor $\kappa_{ij}$ can also be written
as $\kappa_{ij}=\kappa_0 \kappa_{ij}^*(\alpha, a^*)$, where $\kappa_{ij}^*$ is a dimensionless function of $a^*$
and $\alpha$. Thus,
\begin{equation}
\label{4.18} T\partial_T\kappa_{ijk\ell}=T\partial_T \kappa_0
\kappa_{ijk\ell}^*(a^*)=(1-q)\kappa_{ijk\ell}-q\kappa_{ijk\ell}
a^*\partial_{a^*}\ln \kappa_{ijk\ell}^*(a^*).
\end{equation}
In addition, the derivative $\partial_TN_{ijk\ell}^{(0)}$ of the
fourth-degree moments of the zeroth-order distribution can be
evaluated as
\begin{equation}
\label{4.19}
T\partial_TN_{ijk\ell}^{(0)}=4\frac{nT^2}{m^2}\left(2-qa^*\partial_{a^*}\right)
N_{ijk\ell}^*(a^*),
\end{equation}
where I have introduced the reduced fourth-degree moments
\begin{equation}
\label{4.20}
N_{ijk\ell}^*(a^*)=\frac{1}{4}\frac{m^2}{nT^2}N_{ijk\ell}^{(0)}.
\end{equation}
Finally, in dimensionless form, Eq.\ (\ref{4.14}) becomes
\begin{eqnarray}
\label{4.21} \left(\frac{2}{d}a^* P_{xy}^{*}+\zeta^*\right)
\left[1-q(1+a^*\partial_{a^*})\right]\kappa_{ijk\ell}^*&+&\left[\frac{2a^*}{d
}(1-qa^*\partial_{a^*})P_{xy}^*+(q+1)\zeta^*-\nu_{0|3}^*\right]\kappa_{ijk\ell}^*
\nonumber\\
& & - a^*\left(\delta_{ix}\kappa_{jky\ell}^*+
\delta_{jx}\kappa_{iky\ell}^*+\delta_{kx}\kappa_{ijy\ell}^*\right)\nonumber\\
& &
-\frac{\nu_{2|1}^*-\nu_{0|3}^*}{d+2}\left(\delta_{jk}\kappa_{i\ell}^*+
\delta_{ik}\kappa_{j\ell}^*+\delta_{ij}\kappa_{k\ell}^*\right)\nonumber\\
&=&-\frac{4(d-1)}{d(d+2)}\left(2-qa^*\partial_{a^*}\right)
N_{ijk\ell}^*\nonumber\\
& & +\frac{d-1}{d(d+2)}
\left[P_{kj}^{*}(1-qa^*\partial_{a^*})P_{i\ell}^*\right.
\nonumber\\
& & \left. +P_{ik}^{*}(1-qa^*\partial_{a^*})P_{j\ell}^*+
P_{ij}^{*}(1-qa^*\partial_{a^*})P_{k\ell}^*\right], \nonumber\\
\end{eqnarray}
where $\nu_{0|3}^*=\nu_{0|3}/\nu_0$ and
$\nu_{2|1}^*=\nu_{2|1}/\nu_0$.
\begin{figure}
\includegraphics[width=0.6 \columnwidth,angle=0]{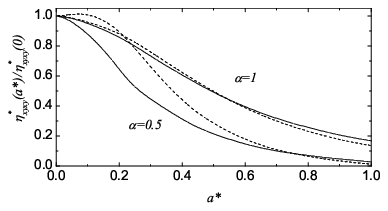}
\caption{Shear-rate dependence of the ratio $\eta_{xyxy}^*(a^*)/\eta_{xyxy}^*(0)$ for two values of the
coefficient of restitution $\alpha$ in the three-dimensional case. The solid lines correspond to the results
obtained for Model A ($q=0$), while the dashed lines refer to the results derived for Model B for
$q=\frac{1}{2}$ by using the approximation (\ref{4.10}). \label{fig4}}
\end{figure}
\begin{figure}
\includegraphics[width=0.6 \columnwidth,angle=0]{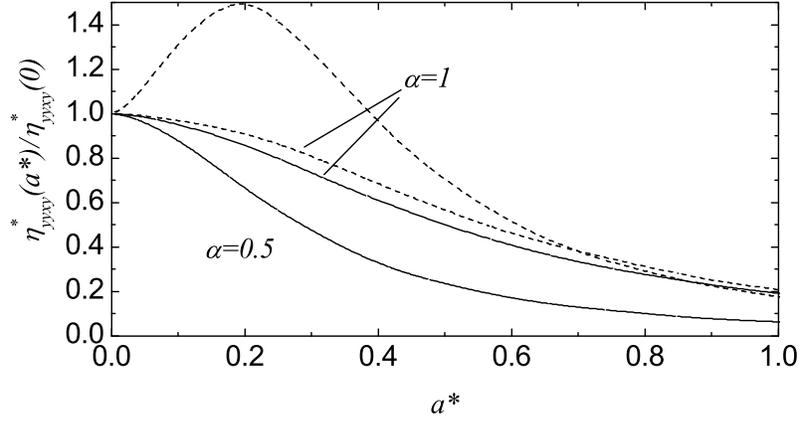}
\caption{Shear-rate dependence of the ratio $\eta_{yyxy}^*(a^*)/\eta_{yyxy}^*(0)$ for two values of the
coefficient of restitution $\alpha$ in the three-dimensional case. The solid lines correspond to the results
obtained for Model A ($q=0$), while the dashed lines refer to the results derived for Model B for
$q=\frac{1}{2}$ by using the approximation (\ref{4.10}). \label{fig5}}
\end{figure}

The set of equations defining the elements of the tensor $\mu_{ijk\ell}$ can be easily obtained by following
similar mathematical steps as those just made for the thermal conductivity tensor. Thus, the reduced tensor
$\mu_{ijk\ell}^*=(n/T\kappa_0)\mu_{ijk\ell}$ is determined from
\begin{eqnarray}
\label{4.22} \left(\frac{2}{d}a^* P_{xy}^{*}+\zeta^*\right)
\left[2-q(1+a^*\partial_{a^*})\right]\mu_{ijk\ell}^*&-&\left(\frac{2a^{*2}}{d
}(\partial_{a^*}P_{xy}^*)-\zeta^*\right)\kappa_{ijk\ell}^*
\nonumber\\
& & -\nu_{0|3}^*\mu_{ijk\ell}^*-
a^*\left(\delta_{ix}\mu_{jky\ell}^*+
\delta_{jx}\mu_{iky\ell}^*+\delta_{kx}\mu_{ijy\ell}^*\right)\nonumber\\
& &
-\frac{\nu_{2|1}^*-\nu_{0|3}^*}{d+2}\left(\delta_{jk}\mu_{i\ell}^*+
\delta_{ik}\mu_{j\ell}^*+\delta_{ij}\mu_{k\ell}^*\right)\nonumber\\
&=&-\frac{4(d-1)}{d(d+2)}\left(1-a^*\partial_{a^*}\right)
N_{ijk\ell}^*\nonumber\\
& & +\frac{d-1}{d(d+2)}
\left[P_{kj}^{*}(1-a^*\partial_{a^*})P_{i\ell}^*\right.
\nonumber\\
& & \left. +P_{ik}^{*}(1-a^*\partial_{a^*})P_{j\ell}^*+
P_{ij}^{*}(1-a^*\partial_{a^*})P_{k\ell}^*\right].\nonumber\\
\end{eqnarray}

In the absence of shear rate ($a^*=0$), the solutions to Eqs.\
(\ref{4.21}) and (\ref{4.22}) give the explicit forms for the
thermal conductivity tensor $\kappa_{ij}^*$ and the tensor
$\mu_{ij}^*$. They can be written as
\begin{equation}
\label{4.23}
\kappa_{ij}^*=\delta_{ij}\frac{d-1}{d}(1+c)\left(\nu_{2|1}^*-2\zeta^*\right)^{-1},
\end{equation}
\begin{equation}
\label{4.24}
\mu_{ij}^*=\frac{\kappa_{ij}^*}{1+c}\frac{\zeta^*+\frac{1}{2}c\nu_{2|1}^*}
{\nu_{2|1}^*-(2-q)\zeta^*},
\end{equation}
where the fourth-cumulant $c$ is \cite{S03}
\begin{equation}
\label{4.25}
c(\alpha)=\frac{12(1-\alpha)^2}{4d-7+3\alpha(2-\alpha)}.
\end{equation}
Equations (\ref{4.23}) and (\ref{4.24}) coincide with the previous
expressions derived by Santos \cite{S03} for the Navier-Stokes
transport coefficients of IMM associated with the heat flux.

 As in the case of the shear viscosity, Eqs.\ (\ref{4.21}) and
(\ref{4.22}) become algebraic for Model A ($q=0$). Even for this model, although the solution to Eqs.\
(\ref{4.21}) and (\ref{4.22}) is simple, it involves a quite tedious algebra due to the presence of the
fourth-degree moments $N_{ijk\ell}^*$ whose expressions for hard spheres ($d=3$) are displayed in Appendix
\ref{appA}. As an illustration, the detailed form of the elements $\kappa_{xy}^*$ and $\kappa_{yy}^*$ of the
thermal conductivity tensor is obtained in Appendix \ref{appC}. A more complete list of the coefficients
$\kappa_{ij}^*$ and $\mu_{ij}^*$ can be obtained from the author upon request.

\section{Comparison with the transport coefficients for IHS}
\label{sec4bis}

As said in the Introduction, the expressions for the generalized transport coefficients of IHS described by the
Bolzmann equation have been recently derived \cite{G06,L06} when the steady state conditions (\ref{2.25}) apply.
These expressions have been obtained by using a BGK-like kinetic model of the Boltzmann equation for a
three-dimensional system \cite{G06}. In this Section, some of the coefficients obtained here for IMM will be
compared with those presented for IHS in the steady state. Beyond the steady state conditions, the dependence of
the generalized transport coefficients $\eta_{ijk\ell}^*$ and $\kappa_{ij}^*$ on both the reduced shear rate
$a^*$ and the coefficient of restitution $\alpha$ will be also analyzed for several situations. In this case and
to the best of my knowledge, there are no available IHS results and so a comparison between IMM and IHS cannot
be carried out.
\begin{figure}
\includegraphics[width=0.45 \columnwidth,angle=0]{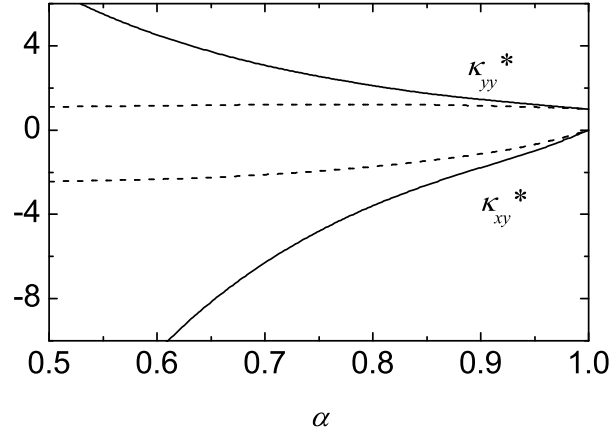}
\caption{Plot of the reduced elements $\kappa_{yy}^*$ and $\kappa_{xy}^*$ of the thermal conductivity tensor as
functions of the coefficient of restitution $\alpha$ in the steady USF state for a three-dimensional system. The
solid lines correspond to IMM for Model A ($q=0$), while the dashed lines refer to the results for IHS derived
in Ref.\ \cite{G06}. \label{fig6}}
\end{figure}
\begin{figure}
\includegraphics[width=0.5 \columnwidth,angle=0]{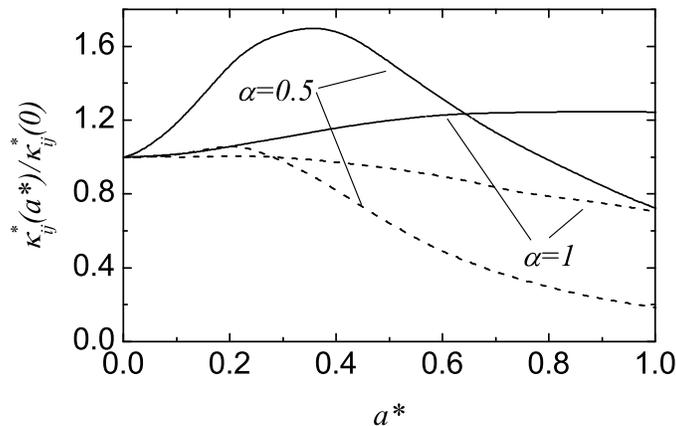}
\caption{Shear-rate dependence of the ratios
$\kappa_{yy}^*(a^*)/\kappa_{yy}^*(0)$ (solid lines) and
$\kappa_{xy}^*(a^*)/\kappa_{xy}^*(0)$ (dashed lines) for two
values of the coefficient of restitution $\alpha$ in the
three-dimensional case. All the results have been derived from
Model A ($q=0$).  \label{fig7}}
\end{figure}

Let us study first the coefficients $\eta_{xyyy}^*$ and $\eta_{yyyy}^*$ in the steady USF state. As said before,
in this situation $a^*$ is related to $\alpha$ through Eq.\ (\ref{2.25}) and so, the coefficient of restitution
is the only control parameter of the system. Figures \ref{fig2} and \ref{fig3} show the $\alpha$-dependence of
$\eta_{xyyy}^*$ and $\eta_{yyyy}^*$, respectively, as given by Model A, Model B with $q=\frac{1}{2}$ by using
the approximation (\ref{4.10}), and IHS \cite{G06} in the steady USF state. It is apparent that the agreement
between the predictions of Model B for IMM (which tries to mimic the true IHS model) and IHS is excellent in the
whole range of values of $\alpha$ analyzed. This confirms the reliability of IMM to reproduce the main trends
observed for the coefficients $\eta_{ijk\ell}^*$. Regarding the results for Models A and B, we observe
quantitative differences between both models for the coefficient $\eta_{yyyy}^*$ for strong dissipation. In the
case of general perturbations, the shear-rate dependence of the ratios $\eta_{yyxy}^*(a^*)/\eta_{yyxy}^*(0)$ and
$\eta_{xyxy}^*(a^*)/\eta_{xyxy}^*(0)$ is plotted in Figs.\ \ref{fig4} and \ref{fig5}, respectively, for two
values of the coefficient of restitution: $\alpha=1$ and $\alpha=0.5$. Here, $\eta_{ijyy}^*(0)$ is the value of
$\eta_{ijyy}^*$ for vanishing shear rates given by Eqs.\ (\ref{4.9.1}) for Model A and Eqs.\ (\ref{4.9.2}) and
(\ref{4.9.3}) for Model B. We observe that, in general, the ratios $\eta_{ijxy}^*(a^*)/\eta_{ijxy}^*(0)$ are
monotonically decreasing functions of the shear rate (shear-thinning effect), except in the region of small
shear rates in the case of Model B for $\eta_{xyxy}^*(a^*)/\eta_{xyxy}^*(0)$ when $\alpha=0.5$. Figures
\ref{fig4} and \ref{fig5} also show that, at a given value of $a^*$, the value of
$\eta_{ijxy}^*(a^*)/\eta_{ijxy}^*(0)$ decrease with dissipation. This means that the inelasticity produces an
inhibition of the momentum transport since the value of $|P_{xy}^{(1)}|$ for inelastic collisions is smaller
than the one obtained in the elastic case.

Next, the thermal conductivity tensor $\kappa_{ij}^*$ is considered when the steady conditions (\ref{2.25})
hold. The dependence of the elements $\kappa_{yy}^*$ and $\kappa_{xy}^*$ on $\alpha$ is plotted in Fig.\
\ref{fig6} for IMM (in the case of Model A) and for IHS \cite{G06}. As happens for the usual thermal
conductivity coefficient \cite{GA05}, the trends observed for IHS are strongly exaggerated by IMM, especially at
high inelasticity. It is possible that the disagreement between both interaction potentials at the level of the
heat flux might be mitigated in part if one considered Model B instead of Model A for IMM. However, this would
require to estimate the fourth-degree moments of the reference state by using the approximation (\ref{3.14}) for
$q=\frac{1}{2}$, which is quite an intricate problem. The shear-rate dependence of $\kappa_{ij}^*$ is
illustrated in Fig.\ \ref{fig7} for the ratios $\kappa_{xy}^*(a^*)/\kappa_{xy}^*(0)$ and
$\kappa_{yy}^*(a^*)/\kappa_{yy}^*(0)$ for two values of the coefficient of restitution: $\alpha=0.5$ and
$\alpha=1$. Here, $\kappa_{xy}^*(0)$ is a Burnett coefficient given by Eq.\ (\ref{c20}) and $\kappa_{yy}^*(0)$
is the Navier--Stokes thermal conductivity coefficient given by Eq.\ (\ref{4.23}) [or, equivalently, Eq.\
(\ref{c21})]. We see that $\kappa_{yy}^*$ increases with $a^*$ in the region of shear rates considered for
elastic collisions, while it does not present a monotonic dependence on $a^*$ in the inelastic case since it
reaches a maximum and then decreases with respect to its Navier--Stokes value. Consequently, while for elastic
gases the shear flow enhances the transport of energy along the direction of the gradient of the flow velocity
($y$ axis), this is not the case for inelastic collisions for large shear rates, where the heat flux is
inhibited by the shearing motion. The off-diagonal element $\kappa_{xy}^*$ measures cross effects in the thermal
conduction since it gives the transport of energy along the $x$ axis due to a thermal gradient parallel to the
$y$ axis. This cross coupling does not appear in the linear regime as the heat flux $q_x^{(1)}$ is at least of
Burnett order (proportional to $a^* \partial T/\partial y$). The element $\kappa_{xy}^*$ is negative and its
absolute value decreases with respect to its Burnett value $\kappa_{xy}^*(0)$ as the shear rate increases,
regardless of the value of dissipation.

\section{Discussion}
\label{sec5}

In this paper, the transport properties of $d$-dimensional IMM in inhomogeneous states close to the USF state
have been analyzed. The physical situation is such that the granular gas is in a state that deviates from the
USF by small spatial gradients. Given that the system is subjected to a strong shear flow, the corresponding
transport coefficients associated with the irreversible heat and momentum fluxes are nonlinear functions of the
shear rate. The explicit evaluation of these coefficients has been the primary objective of this work. The
search for such expressions has been prompted by recent results \cite{G06} derived from a simple kinetic model
of the inelastic Boltzmann equation, which provides explicit expressions for the above coefficients in the case
of IHS. Since this kinetic model is a simplified version of the nonlinear (inelastic) Boltzmann equation, a
still open problem is to get the explicit shear rate dependence of these transport coefficients when the true
Boltzmann collision operator is considered. For this reason, due to the technical difficulties associated with
the mathematical structure of the Boltzmann equation for IHS, the problem studied in Ref.\ \cite{G06} has been
revisited here for IMM. For this interaction model, the collision rate becomes independent of the relative
velocity of the two colliding particles so that the velocity moments of order $k$ of the Boltzmann collision
operator can be {\em exactly} written in terms of moments of order $k'\leq k$ \cite{GS07}. Thanks to this
feature, the second- and fourth-degree velocity moments corresponding to the (pure) USF state has been recently
obtained \cite{SG06}. The knowledge of the latter moments provides the proper basis to explicitly evaluate
transport properties around USF.

As said before, I have been interested in a situation where {\em weak} spatial gradients of density, velocity,
and temperature coexist with a {\em strong} shear rate. Under these conditions, the Boltzmann equation is solved
by means of an extension of the Chapman--Enskog method to arbitrary reference states \cite{L06}. In the case of
the USF state, due to the anisotropy induced by the shear field, tensorial quantities are required to describe
the momentum and heat fluxes instead of the usual Navier--Stokes transport coefficients \cite{single,S03}. In
the first order of the expansion the momentum and heat fluxes are given by Eqs.\ (\ref{3.22}) and (\ref{3.23}),
respectively, where the components of the set of generalized transport coefficient $\eta_{ijk\ell}$,
$\kappa_{ij}$, and $\mu_{ij}$ are the solutions of Eqs.\ (\ref{4.5}), (\ref{4.21}), and (\ref{4.22}),
respectively. Such coefficients are {\em nonlinear} functions of both the shear rate and the coefficient of
restitution. As expected, there are many new transport coefficients in comparison to the case of states near
equilibrium or near the cooling state. These coefficients provide all the information on the physical mechanisms
involved in the transport of momentum and energy under shear flow. Of course, the usual form of the
Navier--Stokes coefficients for IMM is recovered in the absence of shear flow \cite{S03}.

The purpose of this work has been two-fold. First, the evaluation of the shear-rate dependent transport
coefficients of IMM is worthwhile studying by itself as a tractable model to gain some insight into the combined
effect of shear flow and dissipation on the transport properties of the system. Second, the comparison between
the exact results derived here for IMM with those previously obtained for IHS by using a simple kinetic model
allows us to assess the relevance of IMM to reproduce the trends observed in the case of IHS. Recent comparisons
\cite{S03,G03,GA05} carried out between both interaction models show good agreement, especially in the USF
problem \cite{G03}.

The results derived here are given in terms of an effective collision frequency $\omega$, which depends on space
and time through its dependence on the density and temperature. Here, I have considered the general dependence
$\omega \propto n T^q$, where $q=0$ is closer to the original Maxwell molecules for elastic collisions and
$q=\frac{1}{2}$ is closer to hard spheres. To make some contact with the results obtained for IHS \cite{G06},
$\omega$ is chosen to reproduce the cooling rate $\zeta$ of IHS evaluated at the local equilibrium
approximation. This yields the relation (\ref{2.14}). While Model A ($q=0$) lends itself to exact analytical
results for $\eta_{ijk\ell}$, $\kappa_{ij}$, and $\mu_{ij}$, that is not the case for Model B ($q\neq 0$) since
the latter requires to be solved numerically to determine the dependence of the above generalized coefficients
on both the shear rate and dissipation. However, some analytical approximate expressions can be obtained by
considering an expansion of the transport properties in powers of the interaction parameter $q$. In this case,
the results obtained for $q=\frac{1}{2}$ have been compared with those obtained for IHS. Figures \ref{fig2},
\ref{fig3}, \ref{fig6}, and \ref{fig7} illustrate this comparison when the steady state conditions (\ref{2.16})
apply locally [and so, $a^*$ and $\alpha$ are related through Eq.\ (\ref{2.25})]. It is apparent that the IMM
predictions compare quite well with IHS in the case of the viscosity tensor $\eta_{ijk\ell}$, although the
discrepancies between both interaction models increase in the case of the thermal conductivity tensor
$\kappa_{ij}$. In this latter case, the IMM results tend to exaggerate the trends observed for IHS.

An application of the results obtained in this paper would be to perform a linear stability analysis of the
hydrodynamic equations with respect to the steady simple shear flow state. This analysis allows one to determine
the hydrodynamic modes for states near USF as well as the conditions for instabilities at long wavelengths.
Previous results for IHS \cite{G06} in the three-dimensional case indicate that the USF is linearly stable when
the perturbations are along the velocity gradient ($y$ direction) only, while it becomes unstable when the
perturbations are along the vorticity direction ($z$ direction) only. The question now is whether these
conclusions are similar in the case of IMM. Another possible direction of study is to extend the analysis made
here for a single gas to the important subject of granular mixtures. Previous works carried out by the author
and co-workers \cite{G03,GA05} for mixtures of IMM have shown the tractability of the Maxwell kinetic theory for
these complex systems and stimulate the performance of this study in the near future.

\acknowledgments

Partial support from the Ministerio de Ciencia y Tecnolog\'{\i}a (Spain) through Grant No. FIS2007--60977 is
acknowledged.

\appendix
\section{Fourth-degree moments in the USF state for Model A}
\label{appA}

In this Appendix, the explicit expressions of the set of the fourth-degree moments of the zeroth-order
distribution $f^{(0)}$ are given in the case of Model A ($q=0$) for a three-dimensional system ($d=3$). As the
set of independent moments, it is convenient to take the (dimensionless) Ikenberry moments \cite{TM80}
\begin{equation}
\label{a1} \{M_{4|0}^*, M_{2|xx}^*, M_{2|yy}^*, M_{0|yyyy}^*,
M_{0|zzzz}^*, M_{2|xy}^*,M_{0|xxxy}^*, M_{0|xyyy}^*\},
\end{equation}
where
\begin{equation}
\label{a2} M_{4|0}^*=\frac{1}{n}\left(\frac{m}{2T}\right)^2\int\;
d{\bf c} \;c^4 f^{(0)}({\bf c}),
\end{equation}
\begin{equation}
\label{a2bis}
M_{2|ij}^*=\frac{1}{n}\left(\frac{m}{2T}\right)^2\int\; d{\bf c}
\;c^2\left(c_ic_j-\frac{1}{3}c^2\delta_{ij}\right) f^{(0)}({\bf
c}),
\end{equation}
\begin{eqnarray}
\label{a3}
M_{0|ijk\ell}^*&=&\frac{1}{n}\left(\frac{m}{2T}\right)^2\int d{\bf
c}  \left[ c_ic_jc_kc_\ell-\frac{1}{7}c^2
\left(c_ic_j\delta_{k\ell}+c_ic_k\delta_{j\ell}+c_ic_\ell\delta_{jk}
+c_jc_k\delta_{i\ell}+c_jc_\ell\delta_{ik}+c_kc_\ell\delta_{ij}\right)\right.\nonumber\\
&+& \left.\frac{1}{35}c^4\left(\delta_{ij}\delta_{k\ell}+
\delta_{ik}\delta_{j\ell}+\delta_{i\ell}\delta_{jk}\right)\right]f^{(0)}({\bf
c}).
\end{eqnarray}
By using matrix form, the moments (\ref{a1}) are given by \cite{SG06}
\begin{equation}
\boldsymbol{\mathcal{M}}=\boldsymbol{\mathcal{L}}^{-1}\cdot \boldsymbol{\mathcal{C}}, \label{a4}
\end{equation}
where $\boldsymbol{\mathcal{M}}$ is the column matrix defined by the set (\ref{a1})
\begin{equation}
\boldsymbol{\mathcal{M}}=\left(
\begin{array}{c}
M_{4|0}^*\\M_{2|xx}^*\\M_{2|yy}^*\\M_{0|yyyy}^*\\M_{0|zzzz}^*\\M_{2|xy}^*\\M_{0|xxxy}^*
\\M_{0|xyyy}^*,
\end{array}
\right), \label{a5}
\end{equation}
 and $\boldsymbol{\mathcal{L}}$ is the square matrix
\begin{equation}
\boldsymbol{\mathcal{L}}=4\omega_{0|2}^*\gamma
\boldsymbol{\mathcal{I}}+ \boldsymbol{\mathcal{L}}', \label{a6}
\end{equation}
where  $\boldsymbol{\mathcal{I}}$ is the $8\times 8$ identity
matrix and
\begin{equation}
\boldsymbol{\mathcal{L}}'=\left(
\begin{array}{cccccccc}
\omega_{4|0}^*&0&0&0&0&4a^*&0&0\\
0&
\omega_{2|2}^*&0&0&0&\frac{32}{21}a^*&2a^*&0\\
0&0&
\omega_{2|2}^*&0&0&-\frac{10}{21}a^*&0&2a^*\\
0&0&0&
\omega_{0|4}^*&0&-\frac{150}{245}a^*&0&-\frac{12}{7}a^*\\
0&0&0&0&
\omega_{0|4}^*&\frac{24}{245}a^*&\frac{12}{7}a^*&\frac{12}{7}a^*\\
\frac{7}{15}a^*&\frac{2}{7}a^*&\frac{9}{7}a^*&-\frac{7}{3}a^*
&-\frac{1}{3}a^*&
\omega_{2|2}^*&0&0\\
0&\frac{15}{49}a^*&-\frac{6}{49}a^*&-\frac{5}{2}a^*&-\frac{5}{14}a^*&0&
\omega_{0|4}^*&0\\
0&-\frac{6}{49}a^*&\frac{15}{49}a^*&2a^*&\frac{1}{7}a^*&0&0&
\omega_{0|4}^*
\end{array}
\right), \label{a7}
\end{equation}
with \cite{GS07}
\begin{equation}
\label{a8}
\omega_{4|0}^*=\frac{(1+\alpha)^2\left(5+6\alpha-3\alpha^2\right)}{48},
\end{equation}
\begin{equation}
\label{a9}
\omega_{2|2}^*=\frac{(1+\alpha)^2(34+21\alpha-6\alpha^2)}{168},
\end{equation}
\begin{equation}
\label{a9.1}
\omega_{0|4}^*=\frac{(1+\alpha)^2(150+21\alpha-3\alpha^2)}{378}.
\end{equation}

The column matrix $\boldsymbol{\mathcal{C}}$ is
\begin{equation}
\boldsymbol{\mathcal{C}}=\left(
\begin{array}{c}
\mathcal{C}_1\\\mathcal{C}_2\\\mathcal{C}_3\\\mathcal{C}_4\\\mathcal{C}_5
\\\mathcal{C}_6\\\mathcal{C}_7
\\\mathcal{C}_8,
\end{array}
\right), \label{a9.2}
\end{equation}
where
\begin{equation}
\label{a10}
 \mathcal{C}_1= \frac{9}{4}{\lambda}^*_1-
{\lambda}_2^*\left(6{M_{0|yy}^*}^2+2{M_{0|xy}^*}^2\right),
\end{equation}
\begin{equation}
\label{a11}
 \mathcal{C}_2=
\frac{3}{2}{\lambda}^*_3M_{0|xx}^*-\frac{1}{3}
{\lambda}_4^*\left(6{M_{0|yy}^*}^2+{M_{0|xy}^*}^2\right),
\end{equation}
\begin{equation}
\label{a12} \mathcal{C}_3=
\frac{3}{2}{\lambda}^*_3M_{0|yy}^*-\frac{1}{3}
{\lambda}_4^*\left({M_{0|xy}^*}^2-3{M_{0|yy}^*}^2\right),
\end{equation}
\begin{equation}
\label{a13} \mathcal{C}_4=\frac{1}{35}
{\lambda}^*_5\left(81{M_{0|yy}^*}^2-48{M_{0|xy}^*}^2\right),
\end{equation}
\begin{equation}
\label{a14} \mathcal{C}_5=\frac{1}{35}
{\lambda}^*_5\left(81{M_{0|yy}^*}^2+12{M_{0|xy}^*}^2\right),
\end{equation}
\begin{equation}
\label{a15} \mathcal{C}_6= \frac{3}{2}{\lambda}^*_3M_{0|xy}^*+
{\lambda}_4^*M_{0|xy}{M_{0|yy}^*},
\end{equation}
\begin{equation}
\label{a16} \mathcal{C}_7=-\frac{36}{7}
{\lambda}^*_5{M_{0|xy}^*}{M_{0|yy}^*},
\end{equation}
\begin{equation}
\label{a17} \mathcal{C}_8=\frac{27}{7}
{\lambda}^*_5{M_{0|xy}^*}{M_{0|yy}^*}.
\end{equation}
In Eqs.\ (\ref{a10})--(\ref{a17}), I have introduced the second-degree moments
\begin{equation}
\label{a18}
M_{0|ij}^*=\frac{1}{2}\left(P_{ij}^*-\delta_{ij}\right),
\end{equation}
where $P_{ij}^*$ is given by Eqs.\ (\ref{2.20})--(\ref{2.21}) and the quantities
\begin{equation}
\label{a19}
\lambda_1^*=\frac{5}{144}(1+\alpha)^2\left(11-6\alpha+3\alpha^2\right),
\end{equation}
\begin{equation}
\label{a20}
\lambda_2^*=\frac{(1+\alpha)^2\left(1+6\alpha-3\alpha^2\right)}{24},
\end{equation}
\begin{equation}
\label{a21}
\lambda_3^*=\frac{(1+\alpha)^2\left(22-21\alpha+6\alpha^2\right)}{72},
\end{equation}
\begin{equation}
\lambda_4^*=\frac{(1+\alpha)^2\left(21\alpha-1-6\alpha^2\right)}{84},
\end{equation}
\begin{equation}
\label{a22}
\lambda_5^*=\frac{(1+\alpha)^2\left(39-21\alpha+3\alpha^2\right)}{378}.
\end{equation}
The explicit dependence of the fourth-degree moments (\ref{a1}) on $a^*$ and $\alpha$ can be determined from
Eq.\ (\ref{a4}). These expressions are very long and not very illuminating so that they will be omitted here for
the sake of clarity.

The knowledge of the fourth-degree moments of the reference USF
state is needed to compute the thermal conductivity tensor
$\kappa_{ij}$ and the tensor $\mu_{ij}$ [see for instance, Eqs.\
(\ref{4.21}) and (\ref{4.22})]. The relationship between the
(canonical) moments defined in (\ref{a1}) and the moments
$N_{ijk\ell}^*$ defined by Eqs.\ (\ref{4.15.1}) and (\ref{4.20})
is given by
\begin{equation}
\label{a23}
N_{xxxx}^*=\frac{1}{5}M_{4|0}^*+\frac{6}{7}M_{2|xx}^*+\frac{4}{3}(M_{0|yyyy}^*+M_{0|zzzz}^*),
\end{equation}
\begin{equation}
\label{a24}
N_{yyyy}^*=\frac{1}{5}M_{4|0}^*+\frac{6}{7}M_{2|yy}^*+M_{0|yyyy}^*,
\end{equation}
\begin{equation}
\label{a25}
N_{xxyy}^*=\frac{1}{15}M_{4|0}^*+\frac{1}{7}(M_{2|xx}^*+M_{2|yy}^*)-\frac{7}{6}
M_{0|yyyy}^*-\frac{1}{6}M_{0|zzzz}^*,
\end{equation}
\begin{equation}
\label{a26}
N_{xxzz}^*=\frac{1}{15}M_{4|0}^*-\frac{1}{7}M_{2|yy}^*-\frac{1}{6}
M_{0|yyyy}^*-\frac{7}{6}M_{0|zzzz}^*,
\end{equation}
\begin{equation}
\label{a27}
N_{yyzz}^*=\frac{1}{15}M_{4|0}^*-\frac{1}{7}M_{2|xx}^*+\frac{1}{6}
M_{0|yyyy}^*+\frac{1}{6}M_{0|zzzz}^*,
\end{equation}
\begin{equation}
\label{a28} N_{xxxy}^*=\frac{3}{7}M_{2|xy}^*+ M_{0|xxxy}^*,
\end{equation}
\begin{equation}
\label{a29} N_{xyyy}^*=\frac{3}{7}M_{2|xy}^*+ M_{0|xyyy}^*,
\end{equation}
\begin{equation}
\label{a31} N_{xyzz}^*=\frac{1}{7}M_{2|xy}^*-
(M_{0|xxxy}^*+M_{0|xyyy}^*).
\end{equation}

\section{Chapman--Enskog-like expansion}
\label{appB}

In this Appendix, some technical details of the Chapman--Enskog-like type of expansion used in the main text are
described. First, the action of the operators $\partial_t^{(k)}$ on the hydrodynamic fields can be obtained from
the corresponding balance equations associated with the Boltzmann equation (\ref{3.1}). They are given by
\begin{equation}
\label{b1}
\partial_tn+{\bf u}_0\cdot \nabla n=-\nabla \cdot (n\delta {\bf u}),
\end{equation}
\begin{equation}
\label{b2}
\partial_t\delta {\bf u}+{\sf a}\cdot \delta {\bf u}+({\bf u}_0+\delta {\bf u})\cdot \nabla \delta {\bf u}=-
(mn)^{-1}\nabla \cdot {\sf P},
\end{equation}
\begin{equation}
\label{b3} \frac{d}{2}n\partial_tT+\frac{d}{2}n({\bf u}_0+\delta
{\bf u})\cdot \nabla T+aP_{xy}+\nabla \cdot {\bf q}+{\sf P}:\nabla
\delta {\bf u}=-\frac{d}{2}p\zeta,
\end{equation}
where the pressure tensor ${\sf P}$, the heat flux ${\bf q}$ and the cooling rate $\zeta$ are defined by Eqs.\
(\ref{2.10})--(\ref{2.12}), respectively, with the replacement ${\bf V}\rightarrow {\bf c}$. From Eqs.\
(\ref{b1})--(\ref{b3}), it is easy to see that the action of the operator $\partial_t^{(0)}$ is given by Eqs.\
(\ref{3.7}) and (\ref{3.8}), while in the case of the operator $\partial_t^{(1)}$ the result is
\begin{equation}
\label{b4}
\partial_t^{(1)}n+{\bf u}_0\cdot \nabla n=-\nabla \cdot (n\delta {\bf u}),
\end{equation}
\begin{equation}
\label{b5}
\partial_t^{(1)}\delta {\bf u}+({\bf u}_0+\delta {\bf u})\cdot \nabla \delta {\bf u}=
-\frac{1}{m n}\nabla \cdot {\sf P}^{(0)},
\end{equation}
\begin{equation}
\label{b6} \frac{d}{2}n\partial_t^{(1)}T+\frac{d}{2}n({\bf u}_0+
\delta {\bf u})\cdot \nabla T=-aP_{xy}^{(1)}-{\sf P}^{(0)}:\nabla
\delta {\bf u},
\end{equation}
where
\begin{equation}
\label{b7} P_{ij}^{(1)}=\int d{\bf c}\, m c_i c_j  f^{(1)}({\bf
c}).
\end{equation}

Inserting the expansions (\ref{3.3}) and (\ref{3.4}) into Eq.\
(\ref{3.1}), one gets the kinetic equation for the velocity
distribution $f^{(1)}$
\begin{equation}
\label{b8} \left(\partial_t^{(0)}-aV_y\frac{\partial}{\partial
V_x}+{\cal L}\right)f^{(1)}= -\left[\partial_t^{(1)}+({\bf V}+{\bf
u}_0)\cdot \nabla \right]f^{(0)},
\end{equation}
where ${\cal L}$ is the linearized Boltzmann collision operator around the USF defined by Eq.\ (\ref{3.21}). Use
of Eqs.\ (\ref{b4})--(\ref{b6}) in Eq.\ (\ref{b8}) yields
\begin{equation}
\label{b9} \left(\partial_t^{(0)}-aV_y\frac{\partial}{\partial
V_x}+{\cal L}\right) f^{(1)}={\bf Y}_n\cdot \nabla n+{\bf
Y}_T\cdot \nabla T +{\sf Y}_u:\nabla \delta {\bf u},
\end{equation}
where
\begin{equation}
\label{b10} Y_{n,i}=-\frac{\partial f^{(0)}}{\partial
n}c_i+\frac{1}{\rho} \frac{\partial f^{(0)}}{\partial \delta
u_j}\frac{\partial P_{ij}^{(0)}}{\partial n},
\end{equation}
\begin{equation}
\label{b11} Y_{T,i}=-\frac{\partial f^{(0)}}{\partial
T}c_i+\frac{1}{\rho} \frac{\partial f^{(0)}}{\partial \delta
u_j}\frac{\partial P_{ij}^{(0)}}{\partial T},
\end{equation}
\begin{equation}
\label{b12} Y_{u,ij}=n\frac{\partial f^{(0)}}{\partial
n}\delta_{ij}-\frac{\partial f^{(0)}} {\partial \delta
u_i}c_j+\frac{2}{d n}\frac{\partial f^{(0)}}{\partial
T}\left(P_{ij}^{(0)}-a\eta_{xyij}\right).
\end{equation}
Equations (\ref{b9})--(\ref{b12}) are similar to those obtained for IHS \cite{G06}, except that $\zeta^{(1)}=0$
for IMM. The solution to Eq.\ (\ref{b9}) is of the form given by Eq.\ (\ref{3.17}). To get the corresponding
integral equations verifying the unknowns $X_{n,i}$, $X_{T,i}$, and $X_{u,ij}$, one has to take into account the
action of the time derivative $\partial_t^{(0)}$ on the temperature and density gradients,
\begin{eqnarray}
\label{b13}
\partial_t^{(0)} \nabla_i T&=&\nabla_i\partial_t^{(0)}T
\nonumber\\
&=&\left(\frac{2a}{dn^2}(1-n\partial_n) P_{xy}^{(0)}-\frac{\zeta
T}{n}\right) \nabla_i n-\left( \frac{2a}{d n}\partial_T
P_{xy}^{(0)}+(q+1)\zeta\right)\nabla_i T,
\end{eqnarray}
\begin{equation}
\label{b14}
\partial_t^{(0)} \nabla_i \delta u_j=\nabla_i \partial_t^{(0)} \delta u_j=-a_{jk} \nabla_i \delta u_k.
\end{equation}
Substituting the expression of $f^{(1)}$ given by (\ref{3.17})
into (\ref{b8}), the integral equations for $X_{n,i}$, $X_{T,i}$,
and $X_{u,ij}$ are identified as the coefficients of the
independent gradients. This leads to Eqs.\
(\ref{3.18})--(\ref{3.20}).

\section{Transport coefficients}
\label{appC}

In this Appendix, some of the reduced transport coefficients
$\eta_{ijk\ell}^*$, $\kappa_{ij}^*$, and $\mu_{ij}^*$ are
explicitly computed. Let us consider first, the coefficients of
the form $\eta_{ijxy}^*$ for Model A. According to Eqs.\
(\ref{4.7}), these coefficients verify the equations
\begin{equation}
\label{c1}
\left(\frac{4a^*}{d}P_{xy}^*-\omega_{0|2}^*\right)\eta_{xyxy}^*-a^*\eta_{yyxy}^*=
\frac{2}{d}P_{xy}^{*2}-P_{yy}^*,
\end{equation}
\begin{equation}
\label{c2}
\left(\frac{2a^*}{d}P_{xy}^*-\omega_{0|2}^*\right)\eta_{yyxy}^*+
\frac{2a^*}{d}P_{yy}^{*}\eta_{xyxy}^*=\frac{2}{d}P_{xy}^{*}P_{yy}^*,
\end{equation}
where the expressions of the reduced elements $P_{ij}^*$ are given by Eqs.\ (\ref{2.20}) and (\ref{2.21}). Since
$\eta_{yyxy}^*=\eta_{zzxy}^*=\cdots=\eta_{ddxy}^*$ and $\eta_{ijk\ell}^*$ is a traceless tensor, then
$\eta_{xxxy}^*=-(d-1)\eta_{yyxy}^*$. The solution to the set of equations (\ref{c1}) and (\ref{c2}) leads to
Eqs.\ (\ref{4.8}) and (\ref{4.9}) when one takes into account the relations (\ref{2.22}) and (\ref{2.23}). For
small shear rates ($a^*\ll 1$), the coefficients $\eta_{ijxy}^*$ behave as
\begin{equation}
\label{4.9.1} \eta_{xyxy}^*\approx
\frac{1}{\omega_{02}^*}\left(1-\frac{12}{d}\widetilde{a}^2\right),\quad
\eta_{jjxy}^*\approx
\frac{4}{\omega_{02}^*}\frac{1-\delta_{jx}d}{d}\widetilde{a}.
\end{equation}

For Model B, one can estimate $\eta_{ijxy}^*$ by using the
approximation (\ref{4.10}). Neglecting nonlinear terms in $q$, it
is easy to see that the quantities $\Delta \eta_{ijxy}^*$ are
determined from the set of coupled equations
\begin{eqnarray}
\label{c3} \left(\frac{4a^*}{d}P_{xy,A}^*-\omega_{0|2}^*\right)\Delta \eta_{xyxy}^*-a^*\Delta \eta_{yyxy}^*&=&
\frac{4}{d}\left(\Delta P_{xy}^*P_{xy,A}^{*}-a^*\eta_{xyxy,A}^*\Delta
P_{xy}^*\right)-\Delta P_{yy}^*\nonumber\\
& &+\frac{2}{d}\left(a^*P_{xy,A}^*+\zeta^*\right)
(1+a^*\partial_{a^*})\eta_{xyxy,A}^*\nonumber\\
& & -\frac{2}{d}\left(P_{xy,A}^*-a^*\eta_{xyxy,A}^*\right)
a^*\partial_{a^*}P_{xy,A}^*,
\end{eqnarray}
\begin{eqnarray}
\label{c4}
\left(\frac{2a^*}{d}P_{xy,A}^*-\omega_{0|2}^*\right)\Delta
\eta_{yyxy}^*+ \frac{2a^*}{d}P_{yy,A}^{*}\Delta
\eta_{xyxy}^*&=&\frac{2}{d}\Delta
P_{yy}^*\left(P_{xy,A}^{*}-a^*\eta_{xyxy,A}^*\right)\nonumber\\
& &  +\frac{2}{d}\Delta
P_{xy}^*\left(P_{yy,A}^{*}-a^*\eta_{yyxy,A}^*\right)\nonumber\\
& &+\frac{2}{d}\left(a^*P_{xy,A}^*+\zeta^*\right)
(1+a^*\partial_{a^*})\eta_{yyxy,A}^*\nonumber\\
& & -\frac{2}{d}\left(P_{xy,A}^*-a^*\eta_{xyxy,A}^*\right)
a^*\partial_{a^*}P_{yy,A}^*,
\end{eqnarray}
where the elements $\Delta P_{ij}^*$ are given by Eqs.\ (\ref{3.15}) and (\ref{3.16}) and the viscosities
$\eta_{xyxy,A}^*$ and $\eta_{yyxy,A}^*$ for Model A are given by Eqs.\ (\ref{4.8}) and (\ref{4.9}),
respectively. The solution to Eqs.\ (\ref{c3}) and (\ref{c4}), along with the constraint $\Delta
\eta_{xxxy}^*=-(d-1)\Delta \eta_{yyxy}^*$, provides the explicit form for the terms $\Delta \eta_{jjxy}^*$.
Their nonzero elements can be written as
\begin{equation}
\label{c4.1} \Delta \eta_{xyxy}^*=\frac{2 \Lambda_{xy}}{d\omega_{0|2}^{*2} (1+2\gamma)^2(1+6\gamma)^4},
\end{equation}
\begin{equation}
\label{c4.2} \Delta \eta_{xxxy}^*=-\frac{2\widetilde{a}(1-d)\Lambda_{yy}} {d^2\omega_{0|2}^{*2}
(1+2\gamma)^3(1+6\gamma)^4},
\end{equation}
\begin{equation}
\label{c4.3} \Delta \eta_{yyxy}^*=\Delta \eta_{zzxy}^*=-\frac{2\widetilde{a}\Lambda_{yy}} {d^2\omega_{0|2}^{*2}
(1+2\gamma)^3(1+6\gamma)^4},
\end{equation}
where
\begin{eqnarray}
\label{c4.3} \Lambda_{xy}&=&72d\omega_{0|2}^{*}\gamma^4-12\gamma^3\left[9d\omega_{0|2}^{*}+
2(5+2d)\zeta^*\right]+2\gamma^2\left[2(13-2d)\zeta^*-17d\omega_{0|2}^{*}
\right]\nonumber\\
& & +\gamma\left[3d\omega_{0|2}^{*}+ 2(11+4d)\zeta^*\right]-\zeta^*,
\end{eqnarray}
\begin{eqnarray}
\label{c4.4} \Lambda_{yy}&=&144d\omega_{0|2}^{*}\gamma^3-12\gamma^2\left[2d\omega_{0|2}^{*}+
(14+5d)\zeta^*\right]-4\gamma\left[6d\omega_{0|2}^{*}+
(4+7d)\zeta^*\right]\nonumber\\
&& +(d+10)\zeta^*.
\end{eqnarray}
For small shear rates, the coefficients $\eta_{ijxy}^*$ have the
expansions
\begin{equation}
\label{4.9.2} \eta_{xyxy}^*\approx \frac{1}{\omega_{02}^*}\left(1-2q\frac{\zeta^*}{d\omega_{02}^*}\right) +
\frac{\widetilde{a}^2}{d\omega_{02}^*}\left[2q\frac{2(25+4d)\zeta^*+3d\omega_{02}^*} {d\omega_{02}^*}-12\right],
\end{equation}
\begin{equation}
\label{4.9.3} \eta_{jjxy}^*\approx
\frac{4}{d\omega_{02}^*}(1-\delta_{jx}d)\left(1-q\zeta^*\frac{d+10}{2d\omega_{02}^*}
\right)\widetilde{a}.
\end{equation}

The evaluation of the coefficients $\kappa_{ij}^*$ and
$\mu_{ij}^*$ is more intricate than that of $\eta_{ijk\ell}^*$
since it involves the fourth-degree moments of the zeroth-order
distribution. Here, only the coefficients $\kappa_{yy}^*$ and
$\kappa_{xy}^*$ for Model A in the case of a three-dimensional
system ($d=3$) will be considered. These coefficients are given by
\begin{equation}
\label{c5}
\kappa_{xy}^*=\kappa_{xxxy}^*+\kappa_{yyxy}^*+\kappa_{zzxy}^*,
\end{equation}
\begin{equation}
\label{c6}
\kappa_{yy}^*=\kappa_{xxyy}^*+\kappa_{yyyy}^*+\kappa_{zzyy}^*.
\end{equation}
In the case of Model A ($q=0$), the coefficients
$\kappa_{ijk\ell}^*$ obey the set of coupled equations
\begin{eqnarray}
\label{c7} \left[2\left(\frac{2}{3}a^*
P_{xy}^{*}+\zeta^*\right)-\nu_{0|3}^*\right]\kappa_{ijk\ell}^*&-&
\frac{\nu_{2|1}^*-\nu_{0|3}^*}{5}\left(\delta_{jk}\kappa_{i\ell}^*+
\delta_{ik}\kappa_{j\ell}^*+\delta_{ij}\kappa_{k\ell}^*\right)\nonumber\\
& & - a^*\left(\delta_{ix}\kappa_{jky\ell}^*+
\delta_{jx}\kappa_{iky\ell}^*+\delta_{kx}\kappa_{ijy\ell}^*\right)\nonumber\\
&=&-\frac{2}{15}\left(8 N_{ijk\ell}^*-
P_{kj}^{*}P_{i\ell}^*-P_{ik}^{*}P_{j\ell}^*-
P_{ij}^{*}P_{k\ell}^*\right).
\end{eqnarray}
Equation (\ref{c7}) shows that $\kappa_{yy}^*$ and $\kappa_{xy}^*$ are coupled, so that their calculation
involves the set of coefficients
\begin{equation}
\label{c8} \{\kappa_{xxyy}^*, \kappa_{yyyy}^*, \kappa_{zzyy}^*,
\kappa_{xyyy}^*, \kappa_{xxxy}^*, \kappa_{yyxy}^*,
\kappa_{zzxy}^*\}.
\end{equation}
By using matrix notation, the set of seven coupled algebraic
equations for the unknowns (\ref{c8}) can be written as
\begin{equation}
{\mathcal{P}}_{\sigma\sigma'}{\mathcal{Q}}_{\sigma'}=
{\mathcal{R}}_\sigma,\quad \sigma=1,\cdots,7. \label{c9}
\end{equation}
Here, $\boldsymbol{\mathcal{Q}}$ is the column matrix defined by
the set (\ref{c8}), $\boldsymbol{\mathcal{P}}$ is the square
matrix given by
\begin{equation}
\boldsymbol{\mathcal{P}}=\left(
\begin{array}{ccccccc}
\beta+\phi&\phi&\phi&-2a^*&0&0&0\\
3\phi&
\beta+3\phi&3\phi&0&0&0&0\\
\phi&\phi&
\beta+\phi&0&0&0&0\\
0&-a^*&0&
\beta&\phi&\phi&\phi\\
-3a^*&0&0&0&
\beta+3\phi&3\phi&3\phi\\
0&-a^*&0&0&\phi&
\beta+\phi&\phi\\
0&0&-a^*&0&\phi&\phi& \beta+\phi
\end{array}
\right), \label{c10}
\end{equation}
where
\begin{equation}
\label{c11} \beta \equiv 2\left(\frac{2}{3}a^*
P_{xy}^{*}+\zeta^*\right)-\nu_{0|3}^*=-\frac{1+\alpha}{12}\left[12\gamma(1+\alpha)
+7\alpha+2\right],
\end{equation}
\begin{equation}
\label{c12} \phi\equiv
\frac{\nu_{0|3}^*-\nu_{2|1}^*}{5}=\frac{1}{24}(1+\alpha)^2.
\end{equation}
The column matrix $\boldsymbol{\mathcal{R}}$ is
\begin{equation}
\boldsymbol{\mathcal{R}}=-\frac{2}{15}\left(
\begin{array}{c}
8 N_{xxyy}^*-2 P_{xy}^{*2}-P_{xx}^{*}P_{yy}^*\\8 N_{yyyy}^*-3
P_{yy}^{*2}\\8 N_{yyzz}^*- P_{zz}^{*2}\\8 N_{xyyy}^*-3
P_{yy}^{*}P_{xy}^*\\8 N_{xxxy}^*-3 P_{xx}^{*}P_{xy}^*
\\8 N_{xyyy}^*-3
P_{yy}^{*}P_{xy}^*\\8 N_{xyzz}^*- P_{zz}^{*}P_{xy}^*.
\end{array}
\right), \label{c13}
\end{equation}
The solution to Eq.\ (\ref{c9}) is
\begin{equation}
{\mathcal{Q}}_\sigma=({\mathcal{P}}^{-1})_{\sigma\sigma'}
{\mathcal{R}}_{\sigma'}. \label{c14}
\end{equation}
This relation provides an explicit expression for the coefficients
of the form $\kappa_{ijyy}^*$ and $\kappa_{ijxy}^*$. From these
expressions one can get the elements $\kappa_{xy}^*$ and
$\kappa_{yy}^*$. They can be written as
\begin{equation}
\label{c15} \kappa_{xy}^*=\frac{\Psi_{xy}}{\Delta},\quad
\kappa_{yy}^* =\frac{\Psi_{yy}}{\Delta},
\end{equation}
where
\begin{eqnarray}
\label{c16}
\Psi_{xy}&=&a^*\beta(3\beta+8\phi)\mathcal{R}_1+\frac{a^*}{\beta}\mathcal{R}_2\left[
2a^{*2}(3\beta+2\phi)+\beta^2(\beta-2\phi)\right]+\frac{a^*}{\beta}\mathcal{R}_3\left[
\beta^2(\beta-2\phi)-12a^{*2}\phi\right]\nonumber\\
& & +2\mathcal{R}_4a^{*2}\left(3\beta+8\phi\right)+
\left(\mathcal{R}_5+\mathcal{R}_6+\mathcal{R}_7\right)\left[
\beta^2(\beta+5\phi)+6a^{*2}\phi\right],
\end{eqnarray}
\begin{eqnarray}
\label{c17}
\Psi_{yy}&=&\beta^2(\beta+5\phi)\mathcal{R}_1+\mathcal{R}_2\left[
2a^{*2}(\beta+7\phi)+\beta^2(\beta+5\phi)\right]+\mathcal{R}_3\left[
\beta^2(\beta+5\phi)+4a^{*2}\phi\right]\nonumber\\
& & +2\mathcal{R}_4a^{*}\beta\left(\beta+5\phi\right)-
2\left(\mathcal{R}_5+\mathcal{R}_6+\mathcal{R}_7\right)a^*\beta
\phi,
\end{eqnarray}
\begin{equation}
\label{c18} \Delta=2a^{*2}\phi\left(6\beta+23\phi
\right)+\beta^2\left(\beta^2+10\beta \phi+25 \phi^2\right).
\end{equation}
In the limit of small shear rates ($a^*\to 0$), the coefficients
$\kappa_{xy}^*$ and $\kappa_{yy}^*$ behave as
\begin{equation}
\label{c19} \kappa_{xy}^*\approx \kappa_{xy}^{(1)}(\alpha)
a^*+{\cal O}(a^{*3}),\quad \kappa_{yy}^*\approx
\kappa_{yy}^{(0)}(\alpha)+{\cal O}(a^{*2}),
\end{equation}
where
\begin{equation}
\label{c20}
\kappa_{xy}^{(1)}(\alpha)=\frac{112}{5}\frac{162\alpha^7-585\alpha^6+981\alpha^5-111136\alpha^4
+7272\alpha^3+31267\alpha^2-49855\alpha-9466}{(1-9\alpha)^2(1+\alpha)^3\left[3\alpha(\alpha-2)
-5\right]\left[3\alpha(2\alpha-7)-34\right]},
\end{equation}
\begin{equation}
\label{c21}
\kappa_{yy}^{(0)}(\alpha)=-\frac{16}{1+\alpha}\frac{17+9\alpha(\alpha-2)}{(9\alpha-1)\left[
3\alpha(\alpha-2)-5\right]}.
\end{equation}
The coefficient $\kappa_{xy}^{(1)}(\alpha)$ is a Burnett coefficient while $\kappa_{yy}^{(0)}(\alpha)$ gives the
Navier--Stokes thermal conductivity coefficient for a three-dimensional inelastic gas, Eq.\ (\ref{4.23}). When
$\alpha=1$, one recovers well-known results \cite{CC70} for elastic collisions, namely, $\kappa_{yy}^{(0)}=1$
and $\kappa_{xy}^{(1)}=-\frac{7}{2}$.

\end{document}